\documentclass[12pt]{article}

\usepackage{amsmath, amssymb}
\usepackage{graphicx,psfrag}

\newcommand{\be}{\begin{equation}}
\newcommand{\ee}{\end{equation}}
\newcommand{\ba}{\begin{array}}
\newcommand{\ea}{\end{array}}
\newcommand{\bea}{\begin{eqnarray}}
\newcommand{\eea}{\end{eqnarray}}
\newcommand{\nn}{\nonumber}

\newcommand{\dst}{\displaystyle}
\newcommand{\tst}{\textstyle}
\newcommand{\sst}{\scriptstyle}

\newcommand{\p}{\partial}
\newcommand{\xd}{\not{\!\partial}}
\newcommand{\xp}{\not{\mspace{-4mu}p}}

\newcommand{\pr}{\prime}
\newcommand{\ov}{\overline}
\newcommand{\wh}{\widehat}

\newcommand{\RC}{\mathcal{R}}
\newcommand{\vf}{\varphi}
\newcommand{\ep}{\varepsilon}
\newcommand{\eps}{\epsilon}
\newcommand{\La}{\Lambda}
\newcommand{\la}{\lambda}
\newcommand{\da}{\delta}
\newcommand{\Ga}{\Gamma}
\newcommand{\ga}{\gamma}
\newcommand{\si}{\sigma}

\newcommand{\Abar}{\bar{A}}
\newcommand{\Psibar}{\ov \Psi}
\newcommand{\PsiL}{\Psi_\mathrm{L}{}}
\newcommand{\PsiR}{\Psi_\mathrm{R}{}}
\newcommand{\PsiLbar}{{\ov\Psi}_\mathrm{L}{}}
\newcommand{\PsiRbar}{{\ov\Psi}_\mathrm{R}{}}
\newcommand{\chiL}{\chi\raisebox{-5pt}{\scriptsize L}{}}
\newcommand{\chiLbar}{{\ov\chi}\raisebox{-5pt}{\scriptsize L}{}}
\newcommand{\hi}{\;\hat{\!i}}
\newcommand{\hj}{\;\hat{\!\!j}}

\newcommand{\mean}[1]{\langle#1\rangle}
\newcommand{\spm}{\sqrt{p^2+m^2}}

\newcommand{\intc}{\int\!\!}
\newcommand{\bbb}{\!\!\!}
\newcommand{\bb}{\!\!}
\newcommand{\lab}{\raisebox{3pt}{\ba[t]{c} < \\[-3.2mm] \sim \ea}}
\newcommand{\sgn}{\mathrm{sgn}}

\newcommand{\eref}[1]{eq.~(\ref{#1})}
\newcommand{\bt}{\begin{tabular}}
\newcommand{\et}{\end{tabular}}
\newcommand{\bfi}{\begin{figure}}
\newcommand{\efi}{\end{figure}}

\newcommand{\pg}{\psfrag}

\title{\bf To the Fifth Dimension and Back }
\author{\\[1cm]  
 \large{\bf Raman Sundrum}~\thanks{sundrum@pha.jhu.edu}\\[1cm]
 \it Department of Physics and Astronomy\\ 
 \it The Johns Hopkins University\\
 \it 3400 North Charles Street\\
 \it Baltimore, MD 21218, USA}
\date{}

\begin{document}

\numberwithin{equation}{section}

\maketitle

\begin{abstract}
Introductory lectures on Extra Dimensions delivered
at TASI 2004.
\end{abstract}

\section{Introduction}

There are several significant motivations for studying field theory in 
higher dimensions:
(i) We are explorers of spacetime structure, and 
extra spatial dimensions give rise to one of the few possible extensions of 
relativistic spacetime symmetry. 
(ii) Extra dimensions
 are required in string theory 
as the price for taming the bad high energy behavior of 
 quantum gravity within a weakly coupled framework.
(iii) Extra dimensions
 give rise to qualitatively interesting mechanisms within effective field 
theory that may 
play key roles in our understanding of Nature. 
(iv) Extra dimensions can be a type of ``emergent'' phenomenon, 
as best illustrated by the famous AdS/CFT correspondence.

These lectures are intended to provide an introduction, not to the many 
attempts at realistic extra-dimensional model-building, but rather 
to central qualitative extra-dimensional {\it mechanisms}. It is of course 
hoped that by learning these mechanisms in their simplest and most isolated 
forms, the reader is well-equipped to work through more realistic 
incarnations and combinations, in the literature, or better yet,  
as products of their own invention. (Indeed, to really digest 
 these lectures, the reader must use them to understand some particle 
physics models and issues. The other TASI lectures are a good place to start.)
When any of the examples in the lectures 
yields a cartoon of the real world, or a cartoon solution to real world 
problems, I point this out. 

The lectures are organized as follows. Section 2 gives the basic language 
for dealing with field theory in the presence of extra dimensions, 
``compactified'' in order to hide them at low energies. It is also shown how
particles of different spins in four dimensions can be unified 
within a single higher-dimensional field. Section 3 illustrates
 the ``chirality 
problem'' associated with fermions in higher dimensions. 
Section 4 illustrates the emergence of light scalars from higher dimensional 
theories without fundamental scalars, 
computes quantum corrections to the scalar mass (potential), 
and assesses how natural these light scalars are. 
Section 5 describes how extra dimensional boundaries (and boundary 
conditions) can be derived from extra dimensional spaces without boundary, 
by the procedure of ``orbifolding''. It is shown how the chirality problem 
can thereby be solved. The localization of some fields to the boundaries 
is illustrated. Section 6 describes the matching of the higher dimensional 
couplings to the effective four-dimensional long-distance couplings. 
In Section 7, 
the issue of non-renormalizability of higher-dimensional field theory is 
discussed and the scale at which a UV completion is required is identified.
Higher-dimensional General Relativity is discussed in Section 8, in 
partiucular the 
emergence of extra gauge fields at low energies as well as scalar ``radion'' 
fields (or ``moduli'') 
 describing massless fluctuations in the extra-dimensional geometry.
 Section 9 illustrates how moduli may be stabilized to fix the 
extra-dimensional geometry at low energies. Section 10 describes the 
unsolved Cosmological Constant Problem as well as the less problematic 
issue of having a higher-dimensional cosmological constant. Section 10 
shows that  a  higher dimensional cosmological constant leads to 
``warped'' compactifications, as well as the phenomenon of 
``gravity localization''. Section 11 shows that strongly warped 
compactifications naturally lead to hierarchies in the mass scales appearing 
in the low energy effective four-dimensional description. Section 12 shows 
that 
when warped hierarchies are used to generate the Planck/weak-scale hierarchy, 
the extra-dimensional graviton excitations are much more strongly coupled 
to matter than the massless graviton of nature, making them observable at 
 colliders. Section 13 shows how flavor hierarchies and flavor protection 
can arise naturally in warped compactification, following from a study of 
higher-dimensional fermions. Section 14 studies features of 
gauge theory, including the emergence of light scalars, 
 in warped compactifications.

The TASI lectures of Ref. \cite{csaki} and Ref. \cite{Rothstein}, and
the Cargese lectures of Ref. \cite{Rattazzi}, while 
overlapping with the present lectures, 
also contain complementary topics and discussion. 
The central qualitative omissions in the present lectures are supersymmetry, 
which can combine with extra dimensions in 
interesting ways 
(see the TASI lectures of Refs. \cite{csaki} and \cite{Luty}),
a longer discussion of the connection of extra dimensions 
to string theory \cite{ooguri} \cite{joe}, 
a discussion of fluctuating ``branes'' (see Refs. \cite{csaki}
and \cite{Rattazzi}),
and
the (very illuminating) 
AdS/CFT correspondence between some warped 
extra-dimensional theories and some
purely four-dimensional theories with strong dynamics \cite{maldacena} 
\cite{oz} \cite{rscft}.
Phenomenologically, there is no discussion of the ``Large
Extra Dimensions'' scenario \cite{add}, although these lectures
will equip the reader to easily understand it. 

The references included are meant to be useful and to act as gateways to the 
broader literature. They are not intended to be a complete set. I have taken 
moderate pains to get incidental numbers right in the notes, 
but I am fallible. 
I have taken greater pains to ensure that important numbers, such as 
exponents, are correct.

\section{Compactification and Spin Unification}

Let us start by considering $SU(2)$ Yang-Mills (YM) theory in 
five-dimen\-sional (5D) Minkowski spacetime,\footnote{
Our metric signature convention throughout these lectures is
$(+\,-\,\dots\,-)$.}
in particular all dimensions being infinite in size,
\begin{eqnarray}
S &=& {\rm Tr} \int d^4x \int dx_5 
\left\{- \frac{1}{4} F_{MN} F^{MN} \right\}
\nn \\
&=& {\rm Tr} \int d^4x \int dx_5 
\left\{- \frac{1}{4} F_{\mu \nu} F^{\mu \nu}
- \frac{1}{2}  F_{\mu 5} F^{\mu 5} \right\},
\end{eqnarray}
where $M, N = 0,1,2,3,5$ are 5D indices, while $\mu, \nu = 0, 1, 2, 3$ are 
4D indices.
We use matrix notation for $SU(2)$ 
so that the gauge field is $A_M \equiv A_M^a \tau_a$, where $\tau_a$ are the 
isospin Pauli matrices. We will study this theory in an axial gauge, 
$A_5 = 0$.
To see that this is a legitimate gauge, 
imagine that $A_M$ is in a general gauge 
and consider a gauge transformation, 
\begin{equation}
A_M' \equiv \frac{i}{g} \Omega^{-1} D_M \Omega, ~ 
\Omega(x^{\mu}, x_5) \in SU(2),
\end{equation}
where $g$ is the gauge coupling.
It is always possible to find $\Omega(x, x_5)$, such that $A'_5 = 0$.

Ex. Check that this happens for 
$\Omega(x, x_5) = {\mathcal P} e^{i g \int_0^{x_5} dx_5' A_5(x, x_5')}$, 
where ${\mathcal P}$ represents the path-ordering of the exponential.

Ex. Check that in this gauge, 
\begin{equation}
S = {\rm Tr} \int d^4x \int dx_5 
\left\{- \frac{1}{4} F_{\mu \nu} F^{\mu \nu} + 
\frac{1}{2} (\partial_5 A_{\mu})^2 \right\}.
\end{equation}

Let us now compactify the fifth dimension to a circle, 
so that $x_5 \equiv R \phi$, where $R$ is the radius of the circle and $\phi$ 
is an angular coordinate $- \pi \leq \phi \leq \pi$. See Fig. 1. 
\bfi[tb]
\pg{a}{$R$}
\pg{b}{$x^\mu$}
\pg{c}{$x^5\equiv R\phi$}
\begin{center}
\includegraphics{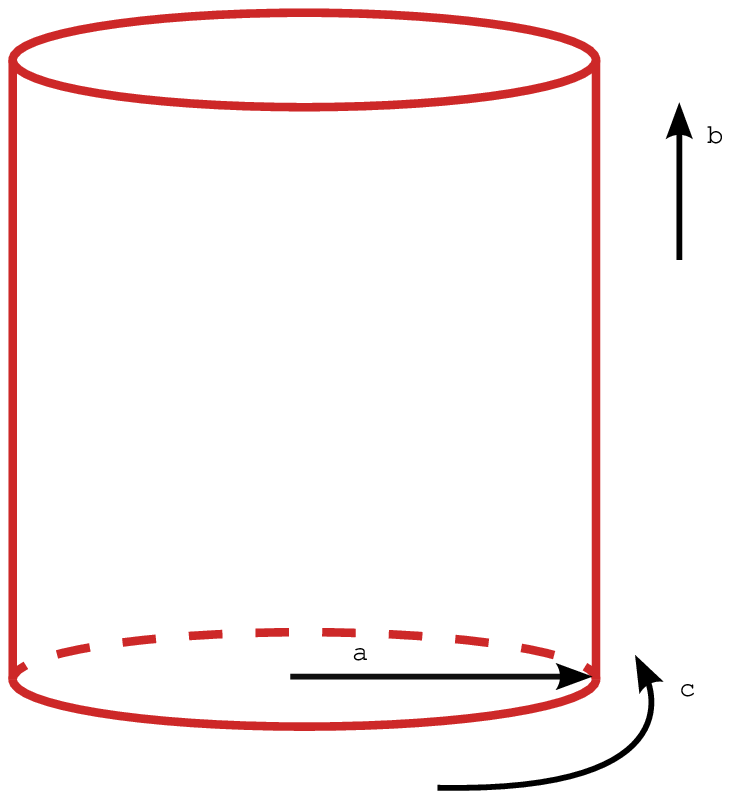}
\end{center}
\caption{A 5D hypercylindrical spacetime}
\efi
We can Fourier expand 
the gauge field in this coordinate, 
\begin{equation}
A_{\mu}(x^{\mu}, \phi) = A_{\mu}^{(0)}(x) + 
\sum_{n=1}^{\infty} (A_{\mu}^{(n)}(x) e^{i n \phi} + {\rm h.c.}).
\end{equation}
But now we can no longer go to axial gauge; in general our $\Omega$ above 
will not be $2 \pi$-periodic. The best we can do is go to an 
``almost axial'' gauge where $A_5$ is $\phi$-independent, 
$A_5(x, \phi) \equiv A_5^{(0)}(x)$, where the action can be written
\begin{eqnarray}
S &=& {\rm Tr} \int d^4x \int^{\pi}_{-\pi} d \phi R 
\left\{- \frac{1}{4} 
F_{\mu \nu} F^{\mu \nu} + \frac{1}{2} (D_{\mu} A_5^{(0)})^2 
+ \frac{1}{2} (\partial_5 A_{\mu})^2 
\right\} \nonumber \\
&=& 2 \pi R \, {\rm Tr} \int d^4x 
\bigg\{ - \frac{1}{2} 
(\partial_{\mu} A_{\nu}^{(0)} 
- \partial_{\nu} A_{\mu}^{(0)})^2
 + \frac{1}{2} (\partial_{\mu} A_5^{(0)})^2 \\
&&+ \sum_{n=1}^{\infty} 
\left[-\frac{1}{2} |\partial_{\mu} A_{\nu}^{(n)} 
- \partial_{\nu} A_{\mu}^{(n)}|^2 + \frac{n^2}{R^2} |A_{\mu}^{(n)}|^2
\right] 
+ {\mathcal O}(A^3)
\bigg\}, \nn
\end{eqnarray}
showing that the 5D theory is equivalent to a 4D theory with an infinite 
tower of 4D fields, with masses, $m_n^2 = n^2/R^2$. This rewriting of 5D
compactified physics is called the Kaluza-Klein (KK) decomposition.

Ex. Show that if $A_M$ is in a general gauge it can be brought to 
almost axial gauge via the (periodic) gauge transformation
\begin{equation}
\Omega(x, \phi) \equiv {\mathcal P} 
e^{i g \int^{\phi}_0 d \phi' R A_5(x, \phi')} 
e^{- i g A_5^{(0)}(x) \phi}.
\end{equation}

Note that the sum over $n$ of the fields in any interaction term must be 
zero since this is just conservation of fifth dimensional momentum, where
 for convenience we define the complex conjugate modes, $A_{\mu}^{(n) *}$, 
to be the modes corresponding to $-n$. In this way a spacetime symmetry 
and conservation law appears as an 
internal symmetry in the 4D KK decomposition, with internal charges, $n$.

Since all of the $n \neq 0$ modes have 4D masses, we can write  
a 4D effective theory valid below  $1/R$ involving just the light 
$A^{(0)}$ modes. Tree level matching yields
\begin{equation}
S_\text{eff} 
{\ba[t]{c} \sim \\[-5pt] \sst E\ll\frac{1}{R} \ea}
2 \pi R\; {\rm Tr} \int d^4x 
\left\{- \frac{1}{4} 
F^{(0)}_{\mu \nu} F^{(0) \mu \nu} + \frac{1}{2} (D_{\mu} A_5^{(0)})^2
\right\}.
\end{equation}
The leading (renormalizable) non-linear interactions  follow entirely from 
the 4D gauge invariance which survives almost axial gauge fixing. 
We have a theory of a 4D gauge field and gauge-charged 4D scalar, 
unified in their higher-dimensional origins. This unification
 is hidden along with 
the extra dimension at low 
energies, but for $E \gg 1/R$ the tell-tale ``Kaluza-Klein'' (KK) 
excitations $A^{(n)}$ are accessible,  and the full story can be 
reconstructed in principle.

Ex. Check that almost axial gauge is preserved by 4D gauge transformations,
$\Omega(x)$ (independent of $\phi$).

Our results are summarized in Fig. 2.
\bfi[tb]
\pg{a}{$n=0$}
\pg{b}{$n=1$}
\pg{c}{$n=2$}
\pg{d}{$n=3$}
\pg{e}{$n=4$}
\pg{f}{\bt{c} 4D \\ effective \\ theory \et}
\pg{m}{$m_{4D}$}
\pg{j}{$J=0$}
\pg{h}{$J=1$}
\pg{k}{\bt{c} complex field \\ 
charged under \\ 5th momentum \et}
\hspace{20pt}
\mbox{
\includegraphics{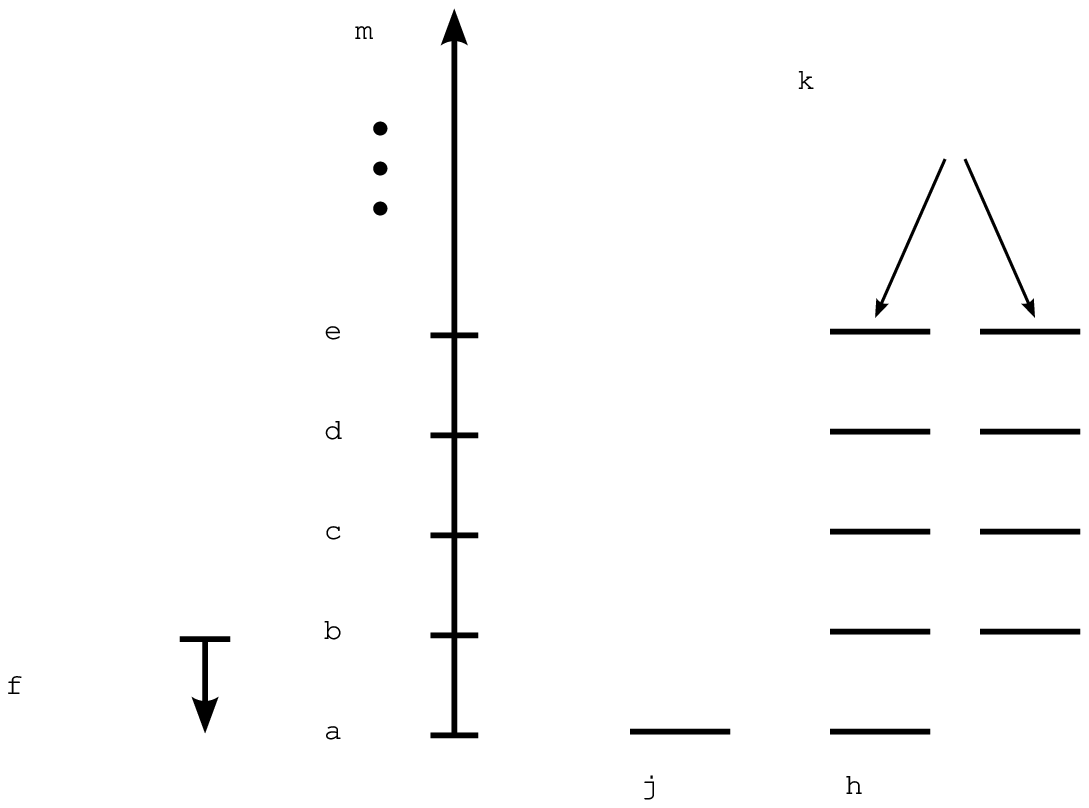}
}
\caption{4D KK spectrum of 5D gauge field}
\efi

\section{5D Fermions and the Chirality Problem} 

To proceed we need a representation of the 5D Clifford algebra, 
$\{\Gamma_M, \Gamma_N \} = 2 \eta_{MN}$. This is straightforwardly provided by 
\begin{equation}
\Gamma_{\mu} \equiv \gamma_{\mu}, ~ ~ \Gamma_5 \equiv - i \gamma_5,
\end{equation}
where the $\gamma$'s are the familiar 4D Dirac matrices. Therefore, 5D fermions
are necessarily $4$-component spinors. We decompose them as 
\begin{equation}
\Psi_{\alpha}(x, \phi) = \sum_{n=- \infty}^{\infty} \Psi_{\alpha}^{(n)}(x) 
e^{i n \phi}. 
\end{equation}
Plugging this into the 5D Dirac action gives
\bea
S_\Psi &=& \intc d^4x\!\!\intc dx_5 \Psibar(i D_M\Ga^M - m)\Psi \nn\\
&=& \intc d^4x\!\!\intc dx_5 \Psibar(i D_\mu\ga^\mu - m)\Psi
-\Psibar\ga_5\p_5\Psi +i g\Psibar A_5\ga_5\Psi \\
&=& 2 \pi R \intc d^4x \sum_{n=-\infty}^{\infty}
\Psibar{}^{(n)}\left(i\ga^\mu\p_\mu-m-i\frac{n}{R}\ga_5\right)\Psi^{(n)}
+ \mathcal{O}(\Psibar A\Psi). \nn
\eea
We see that we get a tower of 4D Dirac fermions labelled by integer $n$ 
(no longer positive), 
with physical masses, 
\be
m_{\mathrm{phys}}^2=m^2+\frac{n^2}{R^2}.
\ee
For small $m$, this is illustrated in Fig. 3.
\bfi[tb]
\pg{a}{$m$}
\pg{b}{$1/R$}
\pg{c}{$2/R$}
\pg{d}{$3/R$}
\pg{e}{$4/R$}
\pg{m}{$m_{4D}$}
\pg{j}{Left}
\pg{h}{Right}
\begin{center}
\includegraphics{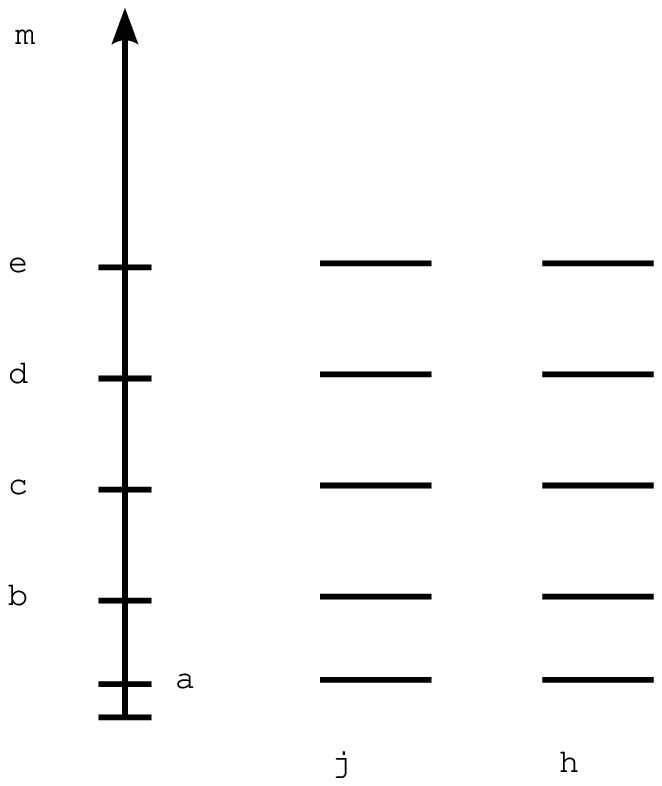}
\end{center}
\caption{4D KK spectrum of 5D fermion}
\efi
These fermions are coupled to the gauge field KK tower, again with all 
interactions conserving 5D momentum, the sum over $n$ of all 4D fields in an 
interaction adding up to zero.

At low energies, $E \ll 1/R$, we can again get a 4D effective action for the 
light $n = 0$ modes,
\bea
S_\text{eff} 
{\ba[t]{c} =\\ 
\sst E\ll\frac{1}{R} \\ 
\sst m\ll\frac{1}{R} \ea}
2 \pi R \intc d^4x 
\left\{
\Psibar{}^{(0)} (i\ga^\mu D_\mu -m)\Psi^{(0)}
+i g\Psibar{}^{(0)}\ga_5 A_5^{(0)}\Psi^{(0)}
\right\} ,
\eea
where the covariant derivative contains only the gauge field $A_{\mu}^{(0)}$.
Note that we also have a Yukawa coupling to the 4D scalar, $A_5^{(0)}$, 
of the same strength as the gauge coupling, so-called gauge-Yukawa 
unification. The idea that the Higgs particle may originate from 
extra-dimensional components of gauge fields was first discussed in 
Refs. \cite{gaugehiggs}.

An  unattractive feature in this cartoon of the real world, emerging below 
$1/R$, is that the 
necessity of having Dirac 4-component spinor representations of 5D Lorentz 
invariance has resulted in having 4-component non-chiral 4D fermion 
zero-modes. The Standard Model is however famously a theory of chiral 
Weyl 2-component fermions. Even as a cartoon this looks worrying.
This general problem in theories of higher dimensions is called the 
``chirality problem'' and we will return to deal with it later.

\section{Light Scalar Quantum Corrections}

Given that light scalars are unnatural in (non-supersymmetric) 
quantum field theories, it is 
rather surprising to see a massless 4D scalar, $A_5^{(0)}$, emerge from 
higher dimensions. Of course, we should consider quantum corrections to our 
classical story and see what happens to the scalar mass. From a purely 4D 
effective field theory viewpoint we would expect there to be large divergent 
corrections to the scalar mass coming from its gauge and Yukawa couplings,
 from diagrams such as Fig. 4, 
\bfi[tb]
\pg{a}{scalar}
\pg{b}{$\Psi$}
\pg{c}{$A_\mu^{(0)}$}
\begin{center}
\includegraphics{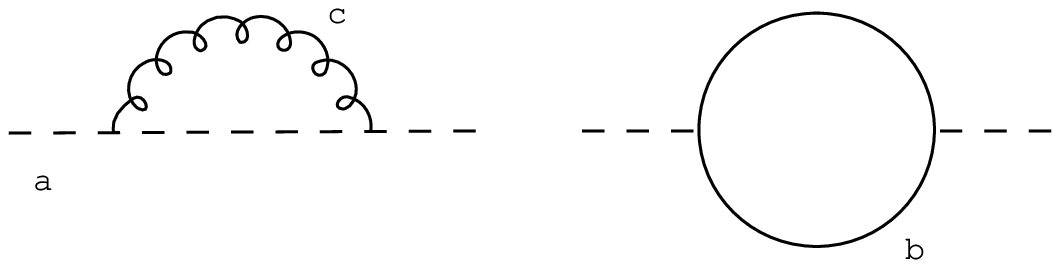}
\end{center}
\caption{Some quantum corrections to 4D scalar self energy}
\efi
%
\be
\da m^2_{\mathrm{scalar}} \sim \frac{g_4^2}{16\pi^2}\La_{\mathrm{UV}}^2\; ,
\ee
suggesting that the scalar is naturally very heavy.
But from the 5D viewpoint $A_5$ is massless because it is part of a 5D gauge 
field, whose mass is protected by 5D gauge invariance. So the question is 
which viewpoint is correct?

To find out let us first compute the 1-fermion-loop effective potential for 
$A_5^{(0)}$ \cite{hosotani}. 
For this purpose we treat $a \equiv g A_5^{(0)}$ as completely 
constant, and $A_{\mu} = 0$. Then,
\be
S_\Psi = 2 \pi R \intc d^4x \sum_n
\Psibar{}^{(n)}\!(x)\left[i\!\!\xd -m 
-i\left(\frac{n}{R}-a\right)\ga_5\right]
\Psi^{(n)}\!(x),
\ee
where 
\be
\xd\equiv\ga^\mu\p_\mu.
\ee
Since $a$ is constant,
\be
S_\Psi =2\pi R \intc \frac{d^4p}{(2\pi)^4} \sum_n
\Psibar{}^{(n)}\!(p)\left[\xp -m
-i\left(\frac{n}{R}-a\right)\ga_5\right]
\Psi^{(n)}\!(p).
\ee
After Wick rotating, this gives
\be
S_\Psi^{\;\mathrm{E}}=\sum_n 2\pi R \intc \frac{d^4p}{(2\pi)^4}\;
\Psibar{}^{(n)}\!(p)\left[\xp+i m
+\left(\frac{n}{R}-a\right)\ga_5\right]
\Psi^{(n)}\!(p).
\ee
Integrating out the fermions by straightforward Gaussian Grassman integration,
\bea
e^{\dst - V_{\mathrm{eff}}} 
&=& \prod_{p,n} \det 
\left[\xp+i m +\left(\frac{n}{R}-a\right)\ga_5\right] \nn\\
&=& \exp\left\{
\sum_n\intc \frac{d^4p}{(2\pi)^4} \;\mathrm{tr}\ln
\left[\xp+i m +\left(\frac{n}{R}-a\right)\ga_5\right]
\right\}.
\eea
From now on, I will simplify slightly by considering a $U(1)$ gauge group 
rather than $SU(2)$. All subtleties will come from finite $R$, so we focus on 
\bea
\frac{\p V_{\mathrm{eff}}}{\p R} 
&=& -\sum_n\intc \frac{d^4p}{(2\pi)^4} \;\mathrm{tr}\left[
-\frac{n}{R^2}\ga_5\frac{1}{\xp+i m+\left(\frac{n}{R}-a\right)\ga_5}
\right] \nn\\
&=& \sum_n\intc \frac{d^4p}{(2\pi)^4} \;\mathrm{tr}\left[
\frac{n}{R^2}\ga_5\frac{\xp-i m+\left(\frac{n}{R}-a\right)\ga_5}
{p^2+\left(\frac{n}{R}-a\right)^2+m^2}
\right] \nn\\
&=& \sum_n\intc \frac{d^4p}{(2\pi)^4} \;\frac{4n(n-a)}{p^2+(n-a)^2+m^2} \; ,
\eea
where we have gone to $R \equiv 1$ units in the last line.  

Naively, this integal and sum over $n$ is quintically divergent! So let us 
carefully regulate the calculation by adding Pauli-Villars fields, in a 5D 
gauge-invariant manner. These fields have the same quantum numbers as 
$\Psi$, but have cut-off size masses $m \rightarrow \Lambda_\text{UV}$, some 
with Bose rather than Fermi statistics. Thereby, 
\bea
\frac{\p V_{\mathrm{eff}}}{\p R}
= \sum_n\intc \frac{d^4p}{(2\pi)^4} \left(
\frac{4n(n-a)}{p^2+(n-a)^2+m^2}
+{\ba{c} \mathrm{regulator}\\ \mathrm{terms} \ea}
\right) < \infty \; .
\eea
The regulator terms resemble the physical term except for having cutoff size 
masses and with signs (determined by the statistics of the regulator field)
chosen in such a way that the entire expression converges. The big trick for 
doing the sum on $n$ is to replace it by a contour integral,
\be
\frac{\p V_{\mathrm{eff}}}{\p R}
= \intc \frac{d^4p}{(2\pi)^4} \oint_C dz \frac{1}{e^{2\pi i z}-1}
\left(\frac{4z(z-a)}{p^2+(z-a)^2+m^2}
+\mathrm{Reg.}
\right),
\ee
where the contour is shown in Fig. 5, 
\bfi[tb]
\pg{a}{$a$}
\pg{b}{$\sst -i\sqrt{p^2+\La_\text{UV}^2}$}
\pg{c}{$\sst -i\sqrt{p^2+m^2}$}
\pg{d}{$\sst i\sqrt{p^2+m^2}$}
\pg{e}{$\sst i\sqrt{p^2+\La_\text{UV}^2}$}
\pg{z}{z}
\pg{k}{C}
\pg{1}{1}
\pg{2}{2}
\pg{3}{3}
\pg{4}{-1}
\pg{5}{-2}
\pg{6}{-3}
\begin{center}
\includegraphics{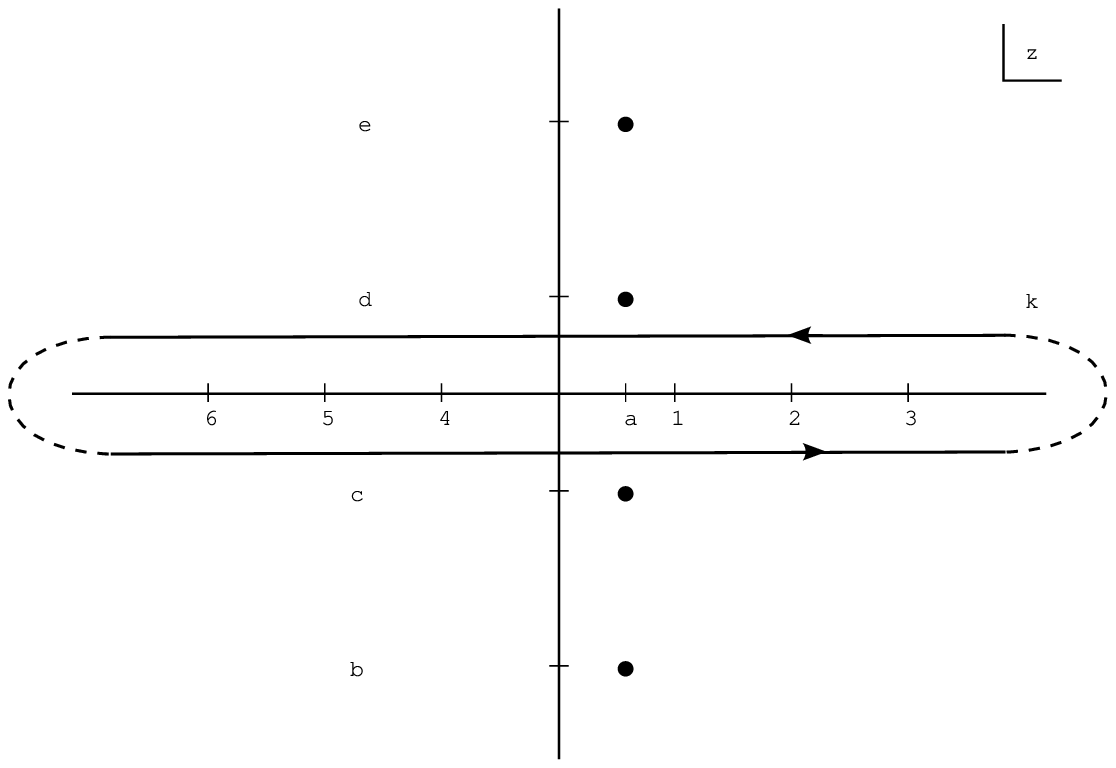}
\end{center}
\caption{Contour for turning KK sum into integral}
\efi
following from the simple poles of 
the factor $1/(e^{2 \pi i z} -1)$ and from the residue theorem. The semi-circles at
infinity needed to have a closed contour are irrelevant because the 
integrand vanishes rapidly enough there, precisely because of the addition of 
the regulator terms.
We can deform 
the contour to that shown in Fig. 6 
\bfi[tb]
\pg{a}{$a$}
\pg{b}{$\sst -i\sqrt{p^2+\La_\text{UV}^2}$}
\pg{c}{$\sst -i\sqrt{p^2+m^2}$}
\pg{d}{$\sst i\sqrt{p^2+m^2}$}
\pg{e}{$\sst i\sqrt{p^2+\La_\text{UV}^2}$}
\pg{k}{$-C_1$}
\pg{n}{$-C_2$}
\pg{z}{z}
\pg{1}{1}
\pg{2}{2}
\pg{3}{3}
\pg{4}{-1}
\pg{5}{-2}
\pg{6}{-3}
\begin{center}
\includegraphics{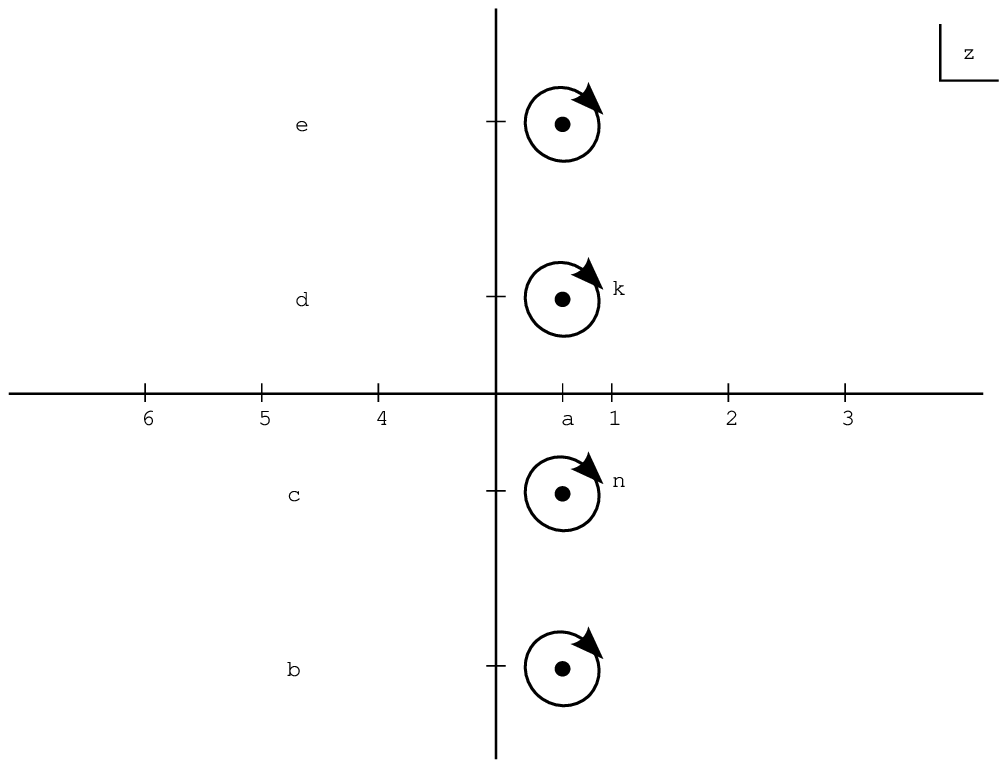}
\end{center}
\caption{Deformed contour for KK sum}
\efi
without encountering any singularities of 
the integrand, so that by the residue theorem,
\bea
\frac{\p V_{\mathrm{eff}}}{\p R}
&=& -4\pi i\intc \frac{d^4p}{(2\pi)^4} \Bigg[
\frac{a+i\sqrt{p^2+m^2}}{e^{2\pi i a}e^{-2\pi\sqrt{p^2+m^2}}-1} \nn\\
&&\hspace{80pt}
+\frac{a-i\sqrt{p^2+m^2}}{e^{2\pi i a}e^{2\pi\sqrt{p^2+m^2}}-1}
+\mathrm{Reg.}\Bigg] . 
\eea
We can also write this as 
\bea
\frac{\p V_{\mathrm{eff}}}{\p R}
&=& 4\pi\intc \frac{d^4p}{(2\pi)^4} \;\Bigg[
\overbrace{
\frac{\sqrt{p^2+m^2}-i a}{e^{2\pi i a}e^{-2\pi\sqrt{p^2+m^2}}-1}
}
-\frac{\sqrt{p^2+m^2}+i a}{e^{2\pi i a}e^{2\pi\sqrt{p^2+m^2}}-1} 
\nn\\
&&\hspace{70pt} 
+(\sqrt{p^2+m^2}-i a)
\left(\frac
{e^{2\pi i a}e^{-2\pi\sqrt{p^2+m^2}}-\overbrace{1}}
{e^{2\pi i a}e^{-2\pi\sqrt{p^2+m^2}}-1}
\right)
\nn\\
&&\hspace{70pt}
-(\sqrt{p^2+m^2}-i a)
+\mathrm{Reg.}
\Bigg],
\eea
where we have just added and subtracted the same quantity in the last two 
terms (not counting the regulator terms). Note that the overbraced terms 
cancel out, leaving 
\bea
\frac{\p V_{\mathrm{eff}}}{\p R}
&=& 4\pi\intc \frac{d^4p}{(2\pi)^4} \;\Bigg[ \left(
-\frac{\sqrt{p^2+m^2}-i a}{e^{-2\pi i a R}e^{2\pi R\sqrt{p^2+m^2}}-1}
\right)
+\mathrm{c.c.} \nn\\
&& \qquad\qquad -\left(\sqrt{p^2+m^2}-i a \right) \Bigg]
+\mathrm{Reg.} 
\eea
where we have put back $R$ explicitly, by dimensional analysis.

Now let us integrate with respect to $R$,
\bea
\label{e29}
V_{\mathrm{eff}}
&=& \intc \frac{d^4p}{(2\pi)^4} \;
\Big\{ -4\mathrm{Re}
\ln \left(1-e^{-2\pi R\sqrt{p^2+m^2}}e^{2\pi i a R} \right)
\nn\\
&&\hspace{20pt}
-4\pi R\left(\sqrt{p^2+m^2}-i a \right)
\Big\} 
+ \;\mathrm{Reg.} \nn\\[2pt]
&&\hspace{20pt}
+ \;\mathrm{irrelevant \;\; const.}
\eea
In the $R \rightarrow \infty$ limit, $V_\text{eff}$ must be independent of $a$ 
since certainly all potential terms for gauge fields vanish by gauge 
invariance as usual. This yields the identity, 
\bea
\label{e30}
V_{\mathrm{eff}}
{\ba[t]{c} 
\longrightarrow \\[-5pt]
\sst R \rightarrow \infty 
\ea}
-4\pi R\intc \frac{d^4p}{(2\pi)^4} \;(\sqrt{p^2+m^2}-i a)
+\mathrm{Reg.}
\equiv \La R \; ,
\eea
where $\Lambda$ is a constant independent of $a$ and $R$. 

Ex. Directly show the cancellation of $a$-dependence in the right hand side 
of \eref{e30} by carefully writing out the regulator terms.
 
Using this identity in \eref{e29} yields 
\bea
V_{\mathrm{eff}}=\La R
-4\intc \frac{d^4p}{(2\pi)^4} \; \mathrm{Re} \ln \left(
1-e^{-2\pi R\sqrt{p^2+m^2}}e^{2\pi i a R} \right)
+\mathrm{Reg.}
\eea

This formula has some remarkable properties. The first term is indeed highly 
cutoff dependent, but it does not depend on $a$. The integrand of the 
second term behaves as $e^{-2 \pi R p}$ for large $p$ and therefore the 
$p$ integrals converge. The regulator terms are suppressed by 
$e^{- 2 \pi R \Lambda_\text{UV}}$ factors and can be completely neglected 
for $\Lambda_\text{UV} R \gg 1$ (or more formally, for $\Lambda_{UV} 
\rightarrow \infty$). We therefore drop the $a$-dependent 
regulator terms from now on.

Finally, combining complex exponentials we arrive at our final result,
\bea
V_{\mathrm{eff}}
&=& \La R
-2\intc \frac{d^4p}{(2\pi)^4} \; \ln \bigg(
1+e^{-4\pi R\sqrt{p^2+m^2}} 
\nn\\ 
&& \qquad\qquad\qquad\qquad
-2\, e^{-2\pi R\sqrt{p^2+m^2}}\cos\left(2\pi R g A_5^{(0)}\right)
\bigg),
\eea
which is illustrated in Fig. 7 .
\bfi[tb]
\pg{a}{$R g A_5^{(0)}$}
\pg{b}{$V_\text{eff}$}
\pg{c}{1}
\pg{d}{$\La R$}
\pg{e}{\bt{c} True\\ vacuum \et}
\begin{center}
\includegraphics{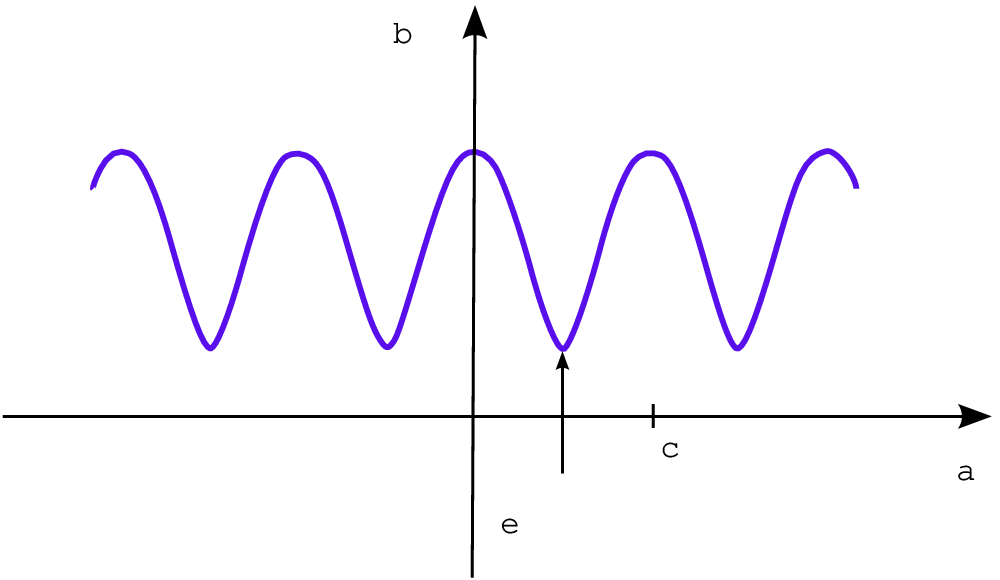}
\end{center}
\caption{Quantum effective potential for $A_5^{(0)}$.
(The potential is not exactly sinusoidal, but it is periodic.)}
\efi
For small $A_5^{(0)}$, this can be approximated,
\bea
V_{\mathrm{eff}} &\sim& \La R
+ \intc \frac{d^4p}{(2\pi)^4} 
\Bigg\{
-4\ln\Bigg(1-e^{-2\pi R\sqrt{p^2+m^2}} \Bigg)
\nn\\[5pt]
&& \mspace{80mu} 
-\left(2\pi R g A_5^{(0)}\right)^2
\left[
\frac{e^{-2\pi R \sqrt{p^2+m^2}}}
{\left(1-e^{-2\pi R \sqrt{p^2+m^2}} \right)^2} 
\right]
\\
&& \mspace{-60mu} 
+\left(2\pi R g A_5^{(0)}\right)^4 \left[
\frac{e^{-2\pi R \sqrt{p^2+m^2}}}
{6\left(1-e^{-2\pi R \sqrt{p^2+m^2}} \right)^2}
+\frac{e^{-4\pi R \sqrt{p^2+m^2}}}
{\left(1-e^{-2\pi R \sqrt{p^2+m^2}} \right)^4} \right]
\Bigg\}. \nn
\eea
We see immediately that the vacuum has non-vanishing  
$\langle A_5^{(0)} \rangle$, 
\be
\left< A_5^{(0)} \right> \sim \frac{1}{R g} \; ,
\ee
for $m < 1/R$.

Let us now return from considering $U(1)$ gauge group back to $SU(2)$. 
Nothing much changes as far as the $\Psi$ loop contribution we have just 
considered ($A_5^{(0)}$ is just to be replaced by
$|A_5^{(0)}| \equiv \sqrt{{\rm tr} A_5^2}$, 
where the trace is over gauged isospin)
 but now there are also diagrams involving gauge loops which 
contribute to the 
$A_5^{(0)}$ effective potential. See Fig. 8.
\bfi[tb]
\pg{a}{$A_5$}
\pg{b}{$A_\mu$}
\pg{c}{$A_\mu^{(n)}$}
\pg{d}{$+$}
\begin{center}
\includegraphics{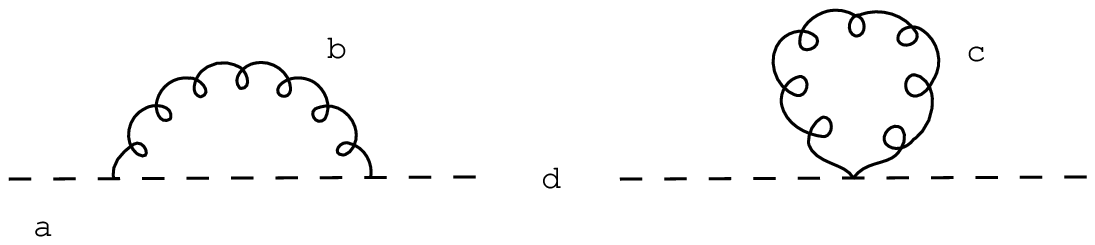}
\end{center}
\caption{Examples of gauge loop contribution to $A_5$ potential}
\efi
By similar methodology, 
these give a contribution illustrated in 
Fig. 9.
\bfi[tb]
\pg{a}{$R g A_5^{(0)}$}
\pg{b}{$V_\text{eff}$}
\begin{center}
\includegraphics{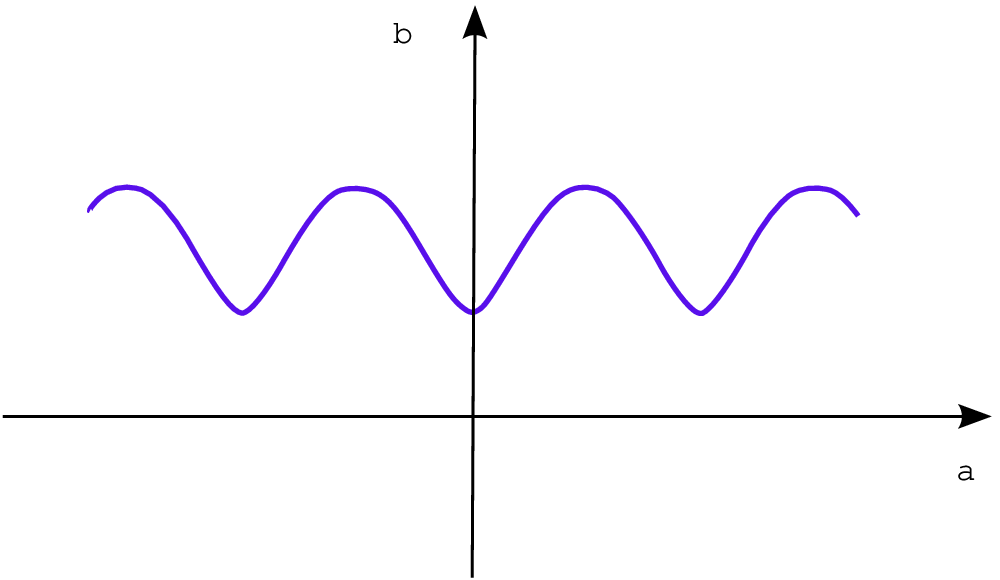}
\end{center}
\caption{Result of gauge loop contribution to $A_5$ potential}
\efi

We see that there is a competition now between the contribution from 
gauge loops which prefers a vacuum at $A_5^{(0)} = 0$ versus the fermion 
loops which prefer a vacuum at $|A_5^{(0)}| \neq 0$. But clearly if we
 include  sufficiently many identical species of $\Psi$, their contribution 
must dominate, and $|A_5^{(0)}| \sim 1/(Rg)$. Since $A_5^{(0)}$ is an 
isovector, a non-zero expectation necessarily breaks the gauge group 
$SU(2)$ down to $U(1)$. One can think of this as a caricature of electroweak 
symmetry breaking where the preserved $U(1)$ is electromagnetism and 
$A_5^{(0)}$ is the Higgs field! 
We refer to it  as ``radiative symmetry breaking'' (also the ``Hosotani 
mechanism'' \cite{hosotani}) because
 it is a loop effect that sculpted out the symmetry breaking 
potential.

In this symmetry breaking vacuum or Higgs phase, we can easily estimate the 
physical mass spectrum, 
\bea
&&m_{\ga^{(0)}} =0 \nn\\[2mm]
&&m_{{W^{\pm}}^{(0)}} \sim \frac{1}{R} \nn\\
&&m_{\Psi^{(0)}} \sim \sqrt{m^2+\frac{1}{R^2}}
{\ba[t]{c} \longrightarrow \\[-5pt] \sst m \,\rightarrow\, 0 \ea}
\frac{1}{R} \nn\\
&&m_{\mathrm{KK}} \sim \frac{1}{R} \nn\\
&&m^2_{\text{``Higgs''}} \sim \frac{g^2}{32\pi^3R^3} \; .
\eea

Now this is certainly an interesting story theoretically, but it is 
surely dangerous to imagine anything like this happening in the real world 
because we are predicting $m_\text{KK} \sim m_W$, and such light KK states 
should already have been seen. However, there is a simple way to make 
the KK scale significantly larger than $m_W$, by making 
\be
\left< A_5^{(0)} \right> \ll \dst\frac{1}{R g} \; .
\ee
Note that for small $A_5^{(0)}$ we have
\bea
V_{\mathrm{eff}} &=&
V_{\mathrm{eff}}^{\Psi\text{--loop}} +
V_{\mathrm{eff}}^{\text{gauge--loop}} 
{\ba[t]{c} \sim \\[-5pt] \sst\text{small}\;a \ea}
\La R \nn\\
&& +\,\left[c_1-c_2(m)N\right](R a)^2
+\left[c_3+c_4(m)N\right](R a)^4 \; ,
\eea
where the $c$'s are order one and positive, and $c_2, c_4$ depend on 
the 5D fermion mass $m$, and $N$ is the number of species of fermions.  
Now let us tune $m$ to achieve
\bea
&&-c_1+c_2(m)N \equiv \ep \ll 1 \nn\\[2pt]
&&\mspace{15mu}c_3+c_4(m)N \sim \mathcal{O}(1) \; ,
\eea
from which it follows that there is a local minimum of the effective 
potential (a possibly cosmologically stable, false vacuum) with 
\be
A_5^{(0)} \sim \dst\frac{\sqrt{\ep}}{g R} \; .
\ee
This yields the hierarchy,
\be
m_{W^{\pm}} \sim 
\frac{\sqrt{\ep}}{R} \sim 
\sqrt{\ep}m_{\mathrm{KK}} \; .
\ee

\section{Orbifolds and Chirality}

If we ask whether our results thusfar could be extended to a realistic 
model of nature, with the standard model as a low energy limit, we encounter
some big problems, not just problems of detail:

a) The previously mentioned chirality problem.

b) Yukawa couplings of the standard model vary greatly. Our low energy 
fermion modes seem to have Yukawa couplings equal to their gauge coupling, 
a reasonable cartoon of the top quark but not of other real world fermions.

A very simple way of solving (a) is to replace the fifth dimensional 
circle by an interval. The two spaces 
can be technically related by realizing the 
interval as an ``orbifold'' of the circle. This is illustrated in Fig. 10,
\bfi[tb]
\pg{a}{$\sst\phi=0$}
\pg{b}[bl][l][1][15]{$\sst\phi=+\pi$}
\pg{c}[bl][l][1][-15]{$\sst\phi=-\pi$}
\pg{d}{\footnotesize \bt{c} 
Identify\\$\sst +\phi$ with $\sst -\phi$\\
(i.e. $\sst x_5$ with $\sst -x_5$) \et}
\pg{e}{$\sst \phi=0$}
\pg{f}{$\sst \phi=\pi$}
\pg{g}{\bt{c} $\sst \phi=0$\\$\sst x_5=0$ \et}
\pg{h}{\bt{c} $\sst \phi=\pi$ \\ $\sst x_5=\pi R$ \et}
\begin{center}
\includegraphics{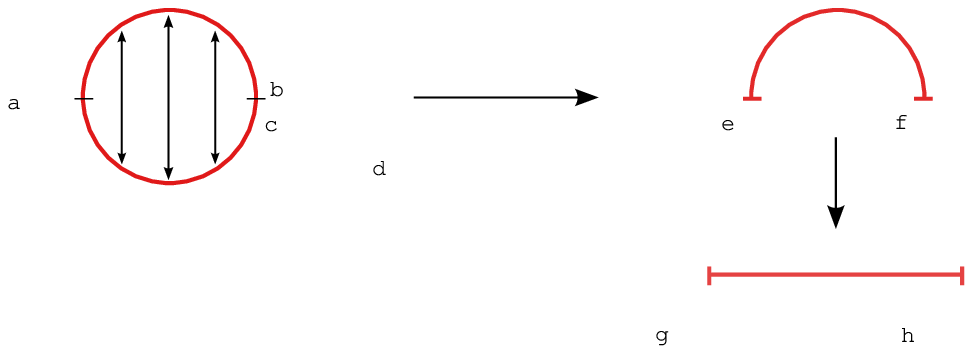}
\end{center}
\caption{Orbifolding the circle to an interval}
\efi
where the points on the two hemispheres of the circle are identified. 
Mathematically, we identify the points at $\phi$ or $x_5$ with 
$-\phi$ or $-x_5$. In this way the physical interval extends a length 
$\pi R$, half the circumference of our original circle. This 
identification is possible if we also assign a ``parity'' transformation 
to all the fields, which is respected by the dynamics (i.e. the action).
The action we have considered above has such a parity, given by 
\be
\ba{ccc}
P(\,x_5)=-x_5 & P(A_\mu)=+A_\mu & P(A_5^{(0)})=-A_5^{(0)} \\[2mm]
  & P(\Psi_\mathrm{L})=+\Psi_\mathrm{L}
  & P(\Psi_\mathrm{R})=-\Psi_\mathrm{R} \; ,
\ea
\ee
precisely when the 5D fermion mass vanishes, $m = 0$. We consider this case 
for now. 

Ex. Check that the action is invariant under this parity transformation.

With such a parity transformation we continue to pretend to live on a circle,
 but with all fields satisfying 
\be
\Phi(x^\mu, -x_5)=P(\Phi)(x^\mu, x_5) \; .
\ee
That is, the degrees of freedom for $x_5 < 0$ are merely a reflection of 
degrees of freedom for $x_5 > 0$, they have no independent existence. Of
 course we also require circular periodicity, 
\be
\Phi(x^\mu, \phi+2\pi)=\Phi(x^\mu, \phi) \; .
\ee
These conditions specify ``orbifold boundary conditions'' on the interval, 
derived from the the circle, which of course has no boundary.

We can write out the mode decompositions (in almost axial gauge) for all the 
fields subject to orbifold boundary conditions, 
\bea
A_\mu(x, \phi) &=& \sum_{n=0}^\infty A_\mu^{(n)}(x)\cos(n\phi) 
\nn\\[3mm]
A_5(x, \phi) &=& 0 \mspace{160mu} \text{Lost ``Higgs''!} 
\nn\\[3mm]
\Psi_\mathrm{L}(x, \phi) &=& \sum_{n=0}^\infty
\Psi_\mathrm{L}^{(n)}(x)\cos(n\phi) 
\nn\\
\Psi_\mathrm{R}(x, \phi) &=& \sum_{n=1}^\infty
\Psi_\mathrm{R}^{(n)}(x)\sin(n\phi)
\qquad \text{Lost} \; \Psi_\mathrm{R}^{(0)} \,\text{!}
\eea
One unfortunate consequence we see is that $A_5$ has no modes, in particular 
orbifolding has eliminated our candidate Higgs!  The good consequence is for 
the chirality problem, in that 
the massless right-handed fermion is eliminated, only the massless left 
handed fermion mode is left. The low energy effective theory below $1/R$ 
is just 
\bea
S_\mathrm{eff} 
{\ba[t]{c} = \\[-5pt] \sst E\ll\frac{1}{R} \ea}
2 \pi R \intc d^4x \left\{
-\frac{1}{4}\left(F_{\mu\nu}^{(0)}\right)^2
+\Psibar{}^{(0)}_\mathrm{L}i D_\mu \ga^\mu \Psi_\mathrm{L}^{(0)}
\right\}.
\eea
With $SU(2)$ gauge group, if $\Psi$ is an isodoublet (so that 
$\Psi_L^{(0)}$ is an isodoublet), the only 
possible gauge invariant mass term for the light mode, 
\be
\Psi_\mathrm{L}{}^i_\alpha \Psi_\mathrm{L}{}^j_\beta
\eps_{ij}\eps^{\alpha\beta} \; ,
\ee
vanishes by fermi statistics. Therefore we apparently 
have a 
chiral effective gauge theory below $1/R$. Unfortunately this theory is 
afflicted by a subtle non-perturbative ``Witten anomaly'', so the theory is 
really unphysical. However, if we consider $\Psi$ to be in the isospin 
$3/2$ representation, we again get a chiral gauge theory, but now not 
anomalous in any way.

Having seen that the chirality problem is soluble, 
we need to recover our 
Higgs field. (For discussion of related mechanisms and further references see 
the TASI review of Ref. \cite{quiros}.) 
To do this we must enlarge our starting gauge group, from 
\be
SU(2) \cong SO(3)
\ee
to $SO(4)$. Gauge fields are conveniently thought of as anti-symmetric 
matrices, $A_M^{ij}$, in the fundmental gauge indices $i, j = 1, 2, 3, 4$.
For simplicity we choose fermions in the fundamental representation, $\Psi^i$.
The action, 
\bea
S &=& {\rm tr} \!\!\intc d^4x \!\!\intc dx_5\Big\{
-\frac{1}{4}F_{\mu\nu}F^{\mu\nu}+\frac{1}{2}(\p_5A_\mu)^2
+\frac{1}{2}(D_\mu A_5^{(0)} )^2 
\nn\\
&& \qquad\qquad\quad
+\;\Psibar i D_\mu \ga^\mu \Psi
-\Psibar\ga_5\p_5\Psi
+i g\Psibar_i A_5^{ij(0)}\ga_5\Psi_j \Big\},
\eea
is invariant under the orbifold parity given by 
\be
\ba{cccc}
P(A_\mu^{\hi\hj})=+A_\mu^{\hi\hj} &
P(A_5^{\hi\hj})=-A_5^{\hi\hj} &
P(\PsiL^{\bbb\hi})=+\PsiL^{\bbb\hi} &
P(\PsiR^{\bbb\hi})=-\PsiR^{\bbb\hi} 
\\[2mm]
P(A_\mu^{\hi 4})=-A_\mu^{\hi 4} &
P(A_5^{\hi 4})=+A_5^{\hi 4} &
P(\PsiL^{\bbb4})=-\PsiL^{\bbb4} &
P(\PsiR^{\bbb4})=+\PsiR^{\bbb4} \; ,
\ea
\ee
where $\hi, \hj = 1, 2, 3$. 

Ex. Check by mode decomposition that this leaves 4D massless fields, 
\be
A_\mu^{\hi\hj(0)},\quad A_5^{\hi 4}{}^{(0)}, \quad
\PsiL^{\bbb\hi}{}^{(0)},\quad \PsiR^{\bbb4}{}^{(0)} \; ,
\ee
that is, a 4D $SO(3)$ gauge field, a 4D Higgs triplet of $SO(3)$, a 
left-handed fermion triplet of $SO(3)$, and a right-handed singlet of 
$SO(3)$. 

This illustrates how (orbifold) boundary conditions on extra 
dimensions can break the gauge group of the bulk of the extra dimensions.
The low-energy effective theory is given by 
\bea
&&S_\mathrm{eff}
{\ba[t]{c} = \\[-5pt] \sst E\ll\frac{1}{R} \ea}
2 \pi R \intc d^4x \Big\{
-\frac{1}{4}F_{\mu\nu}^{(0)}F^{\mu\nu}{}^{(0)}
+\frac{1}{2}(D_\mu A_5^{\hi 4}{}^{(0)} )^2 
\nn\\
&& \mspace{130mu}
+\;\PsiLbar^{\bbb\hi(0)} (i D_\mu \ga^\mu \PsiL^{\bbb(0)})_{\hi}
+\PsiRbar^{\bbb4(0)}i\p_\mu\ga^\mu\PsiR^{\bbb4(0)}
\nn\\[2mm]
&& \mspace{130mu}
+\; i g \left(
\PsiLbar_{\hi}^{\bbb(0)}A_5^{\hi4(0)}\PsiR_4^{\bbb(0)}
+\PsiRbar_4^{\bbb(0)}A_5^{\hi4(0)}\PsiL_{\hi}^{\bbb(0)}
\right)
\Big\}.
\eea
This contains 4D $SO(3)$ gauge theory with two different representations of 
Weyl fermions Yukawa-coupled to a Higgs field. This again bears some 
resemblence to the standard model if we think of the fermion as the left and 
right handed ``top''
quark. But what of the second problem we identified, (b), that the standard 
model contains some fermions with much smaller Yukawa couplings than gauge 
coupling? Such fermions can arise by realizing them very differently in the
higher-dimensional set-up. The simplest example is illustrated in Fig. 11, 
\bfi[tb]
\pg{a}{$x^\mu$}
\pg{b}{\bt{c} $\phi=0$\\$x_5=0$ \et}
\pg{c}{\bt{c} $\phi=\pi$ \\ $x_5=\pi R$ \et}
\pg{d}{$A_M$, $\Psi$}
\pg{e}{$\chi_L(x)$}
\begin{center}
\includegraphics{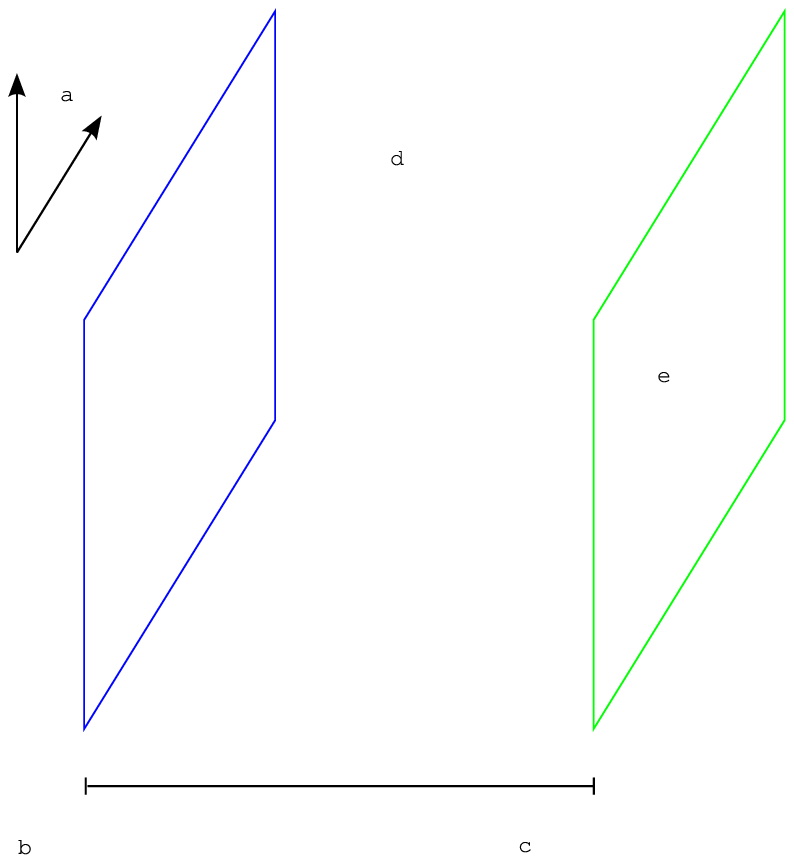}
\end{center}
\caption{Orbifolded higher dimensional spacetime (with boundaries)}
\efi
where beyond the fields we have thusfar considered, which live in the 
``bulk'' of the 5D spacetime, there is a 4D Weyl fermion 
precisely confined to one of the 4D boundaries of the 5D spacetime, 
say $\phi = \pi$. It can couple to the gauge field evaluated at the boundary
 if it carries some non-trivial representation, say triplet. This 
represents a second way in which the chirality problem can be solved, 
localization to a physical 4D subspace or ``3-brane'' (a ``$p$''-brane 
has $p$ spatial dimensions plus time), in this case the boundary of our 5D
spacetime. The new fermion has action, 
\bea
S_\chi =\intc d^4x \;\chiLbar^{\bbb\hi}(x)
\left[i\p_\mu \da^{\hi\hj}+g A_\mu^{\hi\hj}(x,\phi=\pi)\right]
\chiL^{\bbb\hj}(x) \; .
\eea
At low energies, $E \ll 1/R$, this fermion will have identical gauge coupling 
as the $\Psi^{(0)}$ triplet, but it will have no Yukawa coupling, thereby 
giving 
a crude representation of a light fermion of the standard model. 

Well, there are other tricks that one can add to get closer and closer to the 
real world. Ref. \cite{scrucca} gives a nice account of many model-building 
issues and further references. I want to move in a new direction.

\section{Matching 5D to 4D couplings}

Let us study how effective 4D couplings at low energies emerge from the 
starting 5D couplings. Returning to pure $SU(2)$ Yang-Mills on an 
extra-dimensional circle, we get a low-energy 4D theory,
\bea
S_{4\mathrm{eff}}
{\ba[t]{c} \sim \\[-5pt] \sst E\ll\frac{1}{R} \ea}
2 \pi R \intc d^4x \left\{
-\frac{1}{4}F_{\mu\nu}^{(0)}F^{\mu\nu}{}^{(0)}
+\frac{1}{2}(D_\mu A_5^{(0)} )^2 \right\}.
\eea
The fields are clearly not canonically normalized, even though the 5D theory 
we started with was canonically normalized. We can wavefunction renormalize 
the 4D effective fields to canonical form,
\bea
\vf \equiv A_5^{(0)}\sqrt{2\pi R}, \qquad
\Abar_\mu \equiv A_\mu^{(0)}\sqrt{2\pi R} \; ,
\eea
and see what has happened to the couplings, 
\bea
S_{4\mathrm{eff}}
&=& 2 \pi R \intc d^4x \Big\{
-\frac{1}{4}\left(\p_\mu A_\nu^{a(0)}-\p_\nu A_\mu^{a(0)}
-i g_5\eps^{abc}A_\mu^{b(0)}A_\nu^{c(0)}\right)^2 
\nn\\
&&\mspace{140mu}
+\frac{1}{2}\left(
\p_\mu A_5^{(0)}-i g_5 A_\mu^{(0)}A_5^{(0)}\right)^2
\Big\} 
\nn\\
&=& \intc d^4x \Big\{
-\frac{1}{4}\left(\p_\mu \Abar_\nu^a-\p_\nu \Abar_\mu^a
-i \frac{g_5}{\sqrt{2\pi R}}\eps^{abc}\Abar_\mu^b\Abar_\nu^c\right)^2 
\nn\\
&&\mspace{140mu}
+\frac{1}{2}\left(
\p_\mu\vf -i \frac{g_5}{\sqrt{2\pi R}}\Abar_\mu \vf\right)^2
\Big\} .
\eea
From this we read off the effective 4D gauge coupling,
\be
g_{4\mathrm{eff}} = \frac{g_5}{\sqrt{2\pi R}} \; .
\ee

Ex. Check that this is dimensionally correct, 
that  4D gauge couplings are dimensionless
while 5D gauge couplings having units of $1/\sqrt{\rm mass}$.

For experimentally measured gauge couplings, roughly order one, 
we require
\be
g_5 \sim \mathcal{O}(\sqrt{2\pi R}) .
\ee

\section{5D Non-renormalizability}

Now, having couplings with negative mass dimension is the classic sign of 
non-renormalizability, and as you can easily check 
it happens rather readily in higher dimensional quantum field theory. There
 are various beliefs about non-renormalizable theories:

a) A non-renormalizable quantum field theory is an unmitigated disaster. Throw
the theory away at once. Only a few people still hold to this incorrect 
viewpoint. 

b) A non-renormalizable quantum field theory can only be used 
classically, for example in General Relativity where $G_{Newton}$ has 
negative mass dimension. All quantum corrections give nonsense. This 
incorrect view is held by a surprisingly large number of people. 

c) The truth (what I believe): Non-renormalizable theories with couplings, 
$\kappa$, with negative mass dimension can make sense as effective field 
theories, working pertubatively in powers of the dimensionless 
small parameter, 
$\kappa(\rm{Energy})^n$, where $-n$ is the mass dimension of $\kappa$. 
To any {\it fixed} order in this expansion, one in fact has all the 
advantages of  renormalizable quantum field theory. There are 
even meaningful finite 
quantum computations one can perform. In fact we have just 
done one in computing the $A_5$ quantum effective potential. But 
of course there is a price: the whole procedure breaks down once the 
formal small parameter is no longer small, $E \sim 1/\kappa^{1/n}$. 
At higher energies the effective field theory is useless and must be 
replaced by a more fundamental and better behaved description of the 
dynamics. 

Ex. Learn (non-renormalizable) effective field theory at the systematic 
technical level as well as a way of thinking. A good place to start 
is the chiral Lagrangian discussion of soft pions in Ref. \cite{georgi}.

In more detail, perturbative expansions in effective field theory 
will have expansion parameters, $\kappa(\rm{Energy})^n$, divided by 
extra numerical factors such as $2$'s or $\pi$'s. These factors are 
parametrically order one, but enough of them can be quantitatively 
significant. These factors can be estimated from considerations of phase 
space. I will just put these factors in  correctly without explanation.

Ex. Learn the art of naive dimensional analysis, including how to 
estimate  the $2$'s and $\pi$'s (for some discussion in the 
 extra-dimensional context see Ref. \cite{nda}). 
Use this in your work on extra dimensions.

Our findings so far are summarized in Fig. 12. 
\bfi[tb]
\pg{a}{\bt{l} EFT cannot\\ make sense \et}
\pg{b}{5D EFT}
\pg{c}{4D EFT}
\pg{d}{$E$}
\pg{e}{$\La_\text{UV, max} \sim
           \dst\frac{16\pi^2}{g_5^2} \sim
           \dst\frac{8\pi}{R g_4^2}$}
\pg{f}{$\La_{\text{UV}}$}
\pg{g}{$\dst\frac{1}{R} \sim {m_\text{KK}} \sim m_W \sim m_\psi$}
\pg{h}{$m_\text{``Higgs''} \sim 
           \dst\frac{g_5}{\sqrt{32\pi^3 R^3}} \sim 
           \dst\frac{g_4}{4\pi R}$}
\pg{i}{0}
\hspace{50pt}
\mbox{
\includegraphics{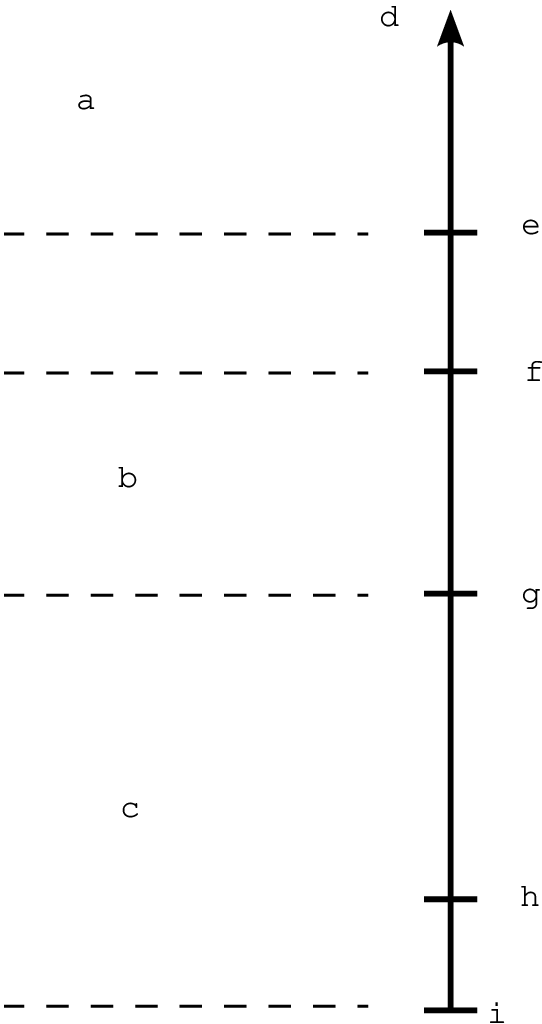}
}
\caption{Natural scales of the 5D gauge theory}
\efi
The non-renormalizable 
effective field theory of 5D gauge theory breaks down when the formal 
small parameter, $E g_5^2$, gets large, that is we can define a 
maximum cutoff on its validity, 
$\Lambda_\text{UV, max} \sim 16 \pi^2/g_5^2$. The 5D 
effective theory cannot hold above this scale and must be replaced by a more 
fundamental theory. Let us say this happens at 
$\Lambda_\text{UV} \leq \Lambda_\text{UV, max}$. 
From here down to $1/R$ we have 5D effective field theory,
 and below $1/R$ we have 4D effective field theory. 

We found it interesting that a Higgs-like candidate emerged from 5D gauge 
fields because it suggested a way of keeping the 4D scalar naturally light, 
namely by identifying it as part of a higher-dimensional vector field. 
But given that  
$\Lambda_\text{UV} \leq 
\Lambda_\text{UV, max} \sim {\mathcal O}(8 \pi/(R g_4^2))$, 
and 4D gauge couplings are 
measured to be not much smaller than one, we must ask how well this 
extra-dimensional picture is doing at addressing the naturalness problem of 
the Higgs. In Fig. 13 
\bfi[tb]
\begin{center}
\bt{c|c}
4D Standard Model &
5D
\\ \hline \\
$\dst\delta_{{}_\text{QM}}m_\text{H}^2 \sim 
\frac{g_4^2+\la_\text{Higgs}}{16\pi^2}\La_\text{UV}^2$ 
&
$\dst m_W \sim g_4 v \sim \frac{1}{R}$
\\[20pt]
\includegraphics[scale=.5]{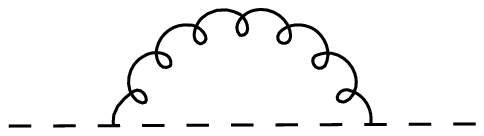}
+
\includegraphics[scale=.5]{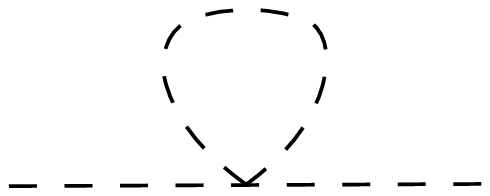}
&
$\dst\La_\text{UV max} \sim \frac{8\pi}{R g_4^2} \sim \frac{8\pi v}{g_4}$
\\[20pt]
$\dst v \sim \frac{m_\text{H}}{\sqrt{\la_\text{Higgs}}} \sim
 \frac{\La_\text{UV}}{4\pi} \sqrt{\frac{g_4^2}{\la_\text{Higgs}}+1} 
\geq \frac{\La_\text{UV}}{4\pi}$
&
$\dst v \sim \frac{g_4}{8\pi} \La_\text{UV max}$
\et
\end{center}
\caption{Natural relationship between $W$ mass and cutoff}
\efi
we make the comparison with purely 4D field theory
with a UV cutoff imposed. We see that in the purely 4D scenario 
one naturally predicts a weak scale 
$v \sim {\mathcal O}(\Lambda_\text{UV}/4 \pi)$, while 
with the 5D regime included, 
$v \sim {\mathcal O}(g_4 \Lambda_\text{UV}/8 \pi)$, 
which is parametrically better. Numerically, it does not look like much 
improvement, and indeed one can do better, but we do not pursue that here.


\section{Gravity-Gauge Unification and the Radion}

The non-renormalizable effective field theory of quantum General Relativity, 
whose small expansion parameter is $E^2 G_{Newton}$, is expected to be 
``UV-completed'' by string theory, just as the Fermi theory of weak 
interactions is completed by the renormalizable electroweak theory. One 
price of calculable string theories is that they predict the existence of 
extra dimensions! You do not have to ask for extra dimensions, they are 
forced on you as the price of maintaining the stringy gauge symmetries 
needed to tame the higher spins in the theory.
 In fact the extra dimensions are required in order to cancel 
quantum anomalies in the stringy gauge symmetries \cite{ooguri}.
However, string theory has been notoriously unpredictive about the
size and shape of the extra dimensions, so we must gedanken experiment, or 
really 
experiment, with the possibilities. Since string theory contains a regime of 
(higher-dimensional) General Relativity, the size and shape of the 
extra dimensions is dynamical as is all of spacetime. This raises several 
interesting issues, which we study below.

Let us begin by returning to 5D spacetime with circular extra dimension. The 
5D gravitational field is given by the metric function on this spacetime, 
in terms of which we can write infinitesimal distances in arbitrary 
coordinates, 
\bea
ds^{\,2} &=& G_{MN}(X)\,dX^M dX^N 
\nn\\[1mm]
&=& G_{MN}(X(X^\pr))\,
\frac{\p X^M}{\p {X^\pr}^{M^\pr}}\,d{X^\pr}^{M^\pr}
\frac{\p X^N}{\p {X^\pr}^{N^\pr}}\,d{X^\pr}^{N^\pr} 
\nn\\[1mm]
&=& G_{M^\pr N^\pr}(X^\pr)\,d{X^\pr}^{M^\pr} d{X^\pr}^{N^\pr} \; .
\eea
From this, we deduce how the metric field transforms under 
5D 
general coordinate transformations (the gauge symmetry 
of 5D General Relativity) in order to maintain 
 the invariance of physical distances,
\be
G_{M^\pr N^\pr}(X^\pr) = G_{MN}(X(X^\pr))
\frac{\p X^M}{\p {X^\pr}^{M^\pr}}
\frac{\p X^N}{\p {X^\pr}^{N^\pr}} \; .
\ee

Expanding around flat (cylindrical) spacetime,
\be
G_{MN}(X)=\eta_{MN}+h_{MN}(X) \; ,
\ee
and working to leading order in $h_{MN}$ and small coordinate transformations,
\be
{X^\pr}^M = X^M-\ep^M(X) \; ,
\ee
we see that the general coordinate transformation of the metric reduces to 
\be
h^\pr_{MN}(X) = h_{MN}(X)+\p_M\ep_N+\p_N\ep_M \; .
\ee
Again, we can go to an almost axial gauge of the form,
\bea
h_{\mu 5}(x,\phi) &=& h_{\mu 5}^{(0)}(x)
\nn\\[2mm]
h_{55}(x,\phi) &=& h_{55}^{(0)}(x)
\nn\\
h_{\mu\nu}(x,\phi) &=& h_{\mu\nu}^{(0)}(x)
+\sum_{n=1}^\infty \left(
h_{\mu\nu}^{(n)}(x)e^{i n\phi} +\text{c.c.}\right),
\eea
where we use coordinates 
\be
X=(x^\mu,\phi) \; .
\ee

Ex. Show how this is to be done using the linearized transformation of the 
metric.

Since the 5D Einstein action, 
\be
{S_{5D}}_\text{\;Einstein}=\intc d^5X
\frac{\sqrt{G}\;\RC}{G_N^{5D}} \; ,
\ee
written in terms of the 5D Ricci scalar curvature ${\mathcal R}$ contains only 
terms with derivatives, the $h_{MN}^{(0)}(x)$ must be massless 4D fields,
without any potential at all. The $h_{MN}^{(n)}$ have $n/R$ masses as usual.
These facts are illustrated in Fig. 14.  
\bfi[tb]
\pg{a}{0}
\pg{b}{$1/R$}
\pg{c}{$2/R$}
\pg{d}{$3/R$}
\pg{e}{$4/R$}
\pg{m}{$m_{4D}$}
\pg{h}{$J=0$}
\pg{i}{$J=1$}
\pg{j}{$J=2$}
\pg{A}{4D GR}
\pg{B}{5D GR}
\begin{center}
\includegraphics{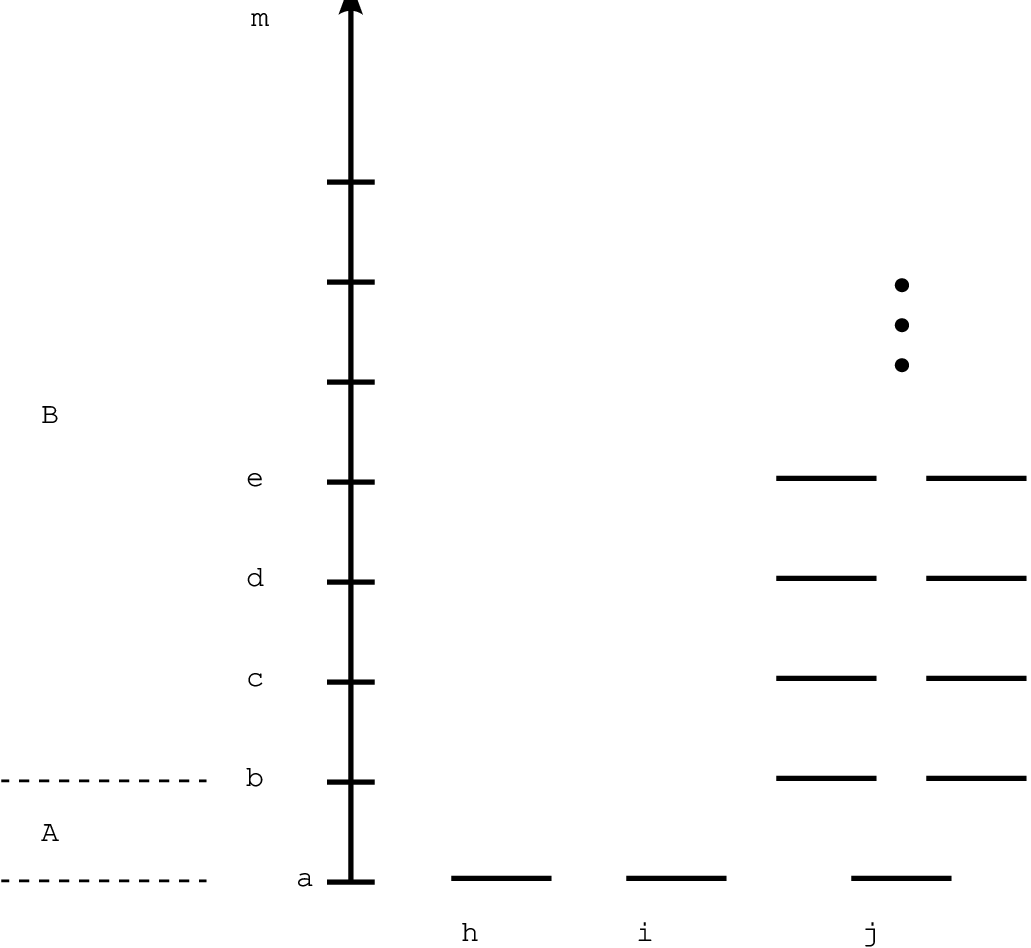}
\end{center}
\caption{KK 4D spectrum of 5D metric decomposition}
\efi
Of course once one goes beyond the 
linearized approximation, all these fields are interacting. The interacting 
massless vector field, $h_{\mu 5}^{(0)}(x)$, must therefore have a protective 
gauge symmetry. 

Ex. Check that in almost axial gauge there is a residual unfixed symmetry,
\be
h_{\mu 5}^\pr(X)=h_{\mu 5}(X)+\p_\mu \ep_5(x) \; .
\ee

Because of this gauge invariance, inherited from 5D general coordinate 
invariance, the quadratic terms involving  $h_{\mu 5}^{(0)}(x)$ in the 
low energy 4D effective action must be just the Maxwell action. This is part 
of the original Kaluza-Klein idea, 4D gauge fields {\it and} 4D General 
Relativity (protected by the residual 4D general coordinate invariance) both 
emerge from a unifed 5D action upon compactification. The 4D effective action 
is tightly constrained by the protective symmetries to be of Maxwell-Einstein 
type, as can be checked by plugging in the massless modes as usual. It is 
easiest to just check this at quadratic order in the action and then use 
the residual symmetries.
Since the vector field 
gauge symmetry corresponds to shifts in the extra dimensional circle (and 
because there is precisely one emergent vector field), the emergent gauge 
group is $U(1)$. Charged matter must correspond to states which 
transform under such extra-dimensional translations. These are precisely 
the KK excitations which carry non-zero extra-dimensional momentum.

It is intriguing to see how ``internal'' gauge groups can 
emerge from the symmetries of the extra-dimensional geometry. One might 
wonder whether more complex and realistic gauge structure, as well as chiral 
4D massless charged matter, can emerge from 
extra-dimensional geometry. While higher dimensional set-ups go in this 
direction, it is in general a very 
difficult game. However, in one way of analysing 
heterotic string theory realistic gauge structure does emerge from several KK 
$U(1)$ gauge fields with special stringy enhancements to non-abelian form 
\cite{joe}. 
In this sense the old Kaluza-Klein program of unification of gauge theory 
(just electromagnetism in the early days) and 
gravity is alive and well.

There is one more 4D massless state which does not follow from a  
residual symmetry, and to that extent is unexpected at first sight, namely 
$h_{55}^{(0)}(x)$. To repeat, it has no potential because the 5D action has 
only derivative terms, and this mode only has $x$-derivatives, being 
independent of the extra dimension by gauge-fixing. This 4D scalar can 
therefore have any VEV, 
\be
\mean{h_{55}^{(0)}}=\xi \; .
\ee
To see what this VEV means physically, note that for the simple VEVs,
\be
\mean{h_{\mu\nu}^{(0)}} =0 \; , \qquad
\mean{h_{\mu 5}^{(0)}} =0 \; ,
\ee
the VEV of the 5D geometry is given by 
\bea
\mean{\,ds^{\,2}} 
&=& \eta_{\mu\nu}dx^\mu dx^\nu -(1+\xi)dx_5 dx_5 
\nn\\
&=& \eta_{\mu\nu}dx^\mu dx^\nu -(1+\xi)R^2 d\phi^2  \; ,
\eea
which is just the geometry of an undeformed (flat) cylinder with 
physical radius $\sqrt{1 + \xi} \; R$.
Thus,  fluctuations in the scalar field correspond to a dynamical radius for 
the extra dimension, as illustrated in Fig. 15. 
\bfi[tb]
\pg{a}{5D}
\pg{b}{$x^\mu$}
\pg{c}{$(G_N^{5D})^{1/3}$}
\begin{center}
\includegraphics{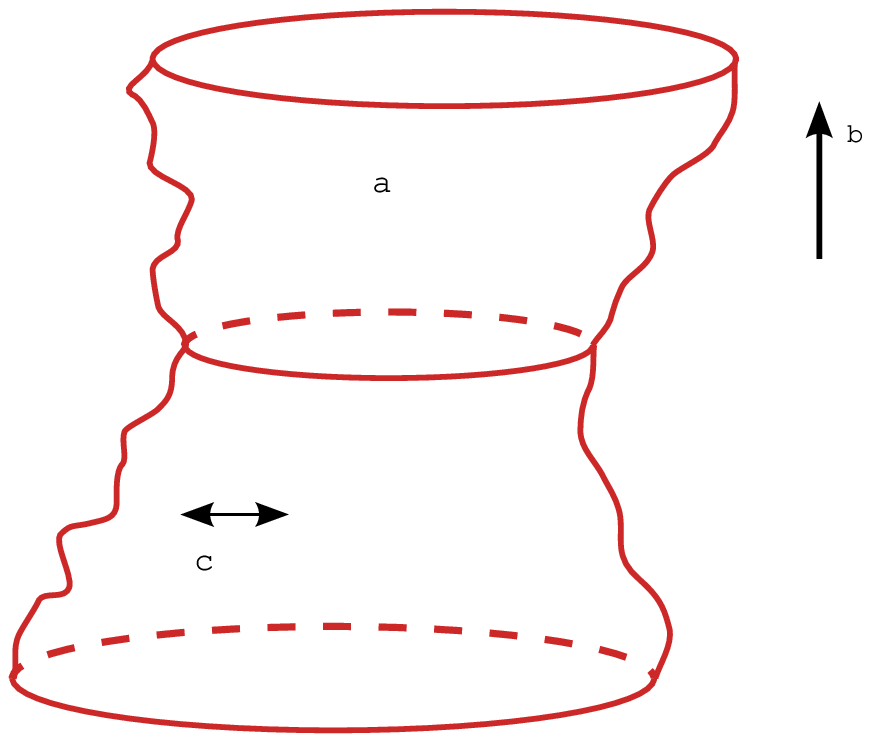}
\end{center}
\caption{Fluctuations of dynamical radius (radion)}
\efi
The quantum of this scalar is therefore referred to as 
the ``radion''.

We are used to scalar fields without potentials in the form of Goldstone 
bosons. But in these cases, any VEV in field space is equivalent to any 
other by the associated symmetry.
The radion field space is an example of a ``moduli space'' of physically 
{\it inequivalent} vacua (differing radii are physically quite different 
obviously). The radion is an example of a ``modulus''. The string theories 
that have been studied start life with very large moduli spaces of 
inequivalent vacua. Thus even a unique theory can lose predictivity in 
the maze of physically inequivalent vacua Nature has to choose from.

The good news is that we see a cartoon of the 4D ingredients needed for
the real world emerging from our very simple unified 5D example: 4D scalars, 
4D gauge theory and 4D gravity. If one throws in supersymmetry, another key 
ingredient of superstring theory, one must necessarily have fermions as well.
The size and shape of the extra dimensions, and the low energy 4D world 
they produce, is determined by the moduli VEVs. Small corrections to 
the vanishing potential of moduli space can actually favor one or a discrete set 
of vacua, yielding greater predictivity (in string theory right now ``greater''
does not mean great). Let us continue with our 5D cylinder to see a simple 
example of such small corrections, in this case from quantum effects related 
to the famous Casimir effect.

\section{Modulus Stabilization}

Let us add a 5D fermion to the (gravitating) cylinder. Our old effective 
potential calculation from fermion loops, 
now dropping $A_5$ since there are no 5D gauge fields, is given by
\be
V_\text{eff} \sim \La R -4\intc \frac{d^4p}{(2\pi)^4} \,\ln \left(
1-e^{-2\pi R\spm} \right).
\ee
In the 5D general relativistic context we should consider the $1$-fermion-loop
induced effective potential for the dynamical radius VEV, $R$. Recall, 
$\Lambda$ contains all UV divergences in this calculation. The renormalized 
value of $\Lambda$ is frequently tuned to zero (more on this later) so that 
crudely,
\be
V_\text{eff} 
{\ba[t]{c} \sim \\[-5pt] \sst m\gg\frac{1}{R} \ea}
e^{-2\pi R m}\int_0^m \!\!d^4p \sim
m^4 e^{-2\pi R m} \; .
\ee
Now, let us assume that there is also a 5D scalar boson as well and add 
its quantum-loop contribution to the $R$ effective potential. In 
complete analogy to our fermion loop computation, except for a sign 
due to the different statistics, 
\be
V_\text{eff} \sim
m^4 e^{-2\pi R m}
-\mu^4 e^{-2\pi R\mu} \; .
\ee
 Upon minimizing the effective potential with respect to the dynamical radius,
\be
e^{\dst 2\pi R(m-\mu)} \sim \left(\frac{m}{\mu}\right)^5, \qquad
R \sim \frac{5 \ln\left(\dst\frac{m}{\mu}\right)}{2\pi(m-\mu)} \; .
\ee
That is the corrections have selected a particular radius from the original 
moduli space!

The general conclusion is that in (non-supersymmetric) extra dimensions 
small corrections can generate an effective potential for moduli which 
stabilizes the size and shape of the extra dimensions. This radius can 
readily be moderately larger than the fundamental length scale of the 
cutoff of non-renormalizable higher-dimensional effective field theory. In 
the present example, this is achieved by taking the matter masses, $m, \mu$,
to be moderately lighter than the mass scale of the cutoff. 

If the present example is orbifolded so that the extra dimension is an
interval,
one can also add a constant "potential" to be localized to one of the 
boundaries. Since such a localized term is insensitive to $R$, it will just 
contribute an $R$-independent constant to the effective potential. 
Such a constant seems irrelevant from  the point of view of the radion, but 
it is important when the effective dynamics of 4D General Relativity is also 
taken into account, as we discuss in the next section.

\section{The Cosmological Constant (Problem)} 

The dominant interactions in 4D General Relativity at long distances include 
not just the Einstein action but also a cosmological constant term,
\be
S=\intc d^4x \sqrt{-g}
\left\{- \la +\frac{\RC}{16\pi G_N} 
\right\}.
\ee
In flat space 
quantum matter generates zero-point energies or vacuum energy, 
corrected by interactions in all possible ways, but we usually throw this away 
as a physically irrelevant field-independent constant in the effective action. 
But when coupled to gravity, all vacuum energy from matter and radiation 
contributes to the cosmological term. If we trust the standard model up to 
TeV energies, one naturally expects contributions to $\lambda$ of order 
at least TeV$^4$. 
Crudely, in the presence of such a term Einstein's Equations read
\be
\RC \sim G_N \la \; ,
\ee
where the left-hand side is a measure of the curvature of spacetime. In 
particular, flat Minkowski spacetime is not a solution to Einstein's 
equations for $\lambda \neq 0$. If $\lambda \sim $ TeV$^4$, the 
radius of curvature of spacetime would be 
\be
\frac{1}{\sqrt{\RC}} \sim 
\frac{M_{Pl}}{\sqrt{\la}} \sim
\frac{M_{Pl}}{\text{TeV}^2} \sim
\text{1mm !}
\ee
This would be tremendously at odds with even everyday experience, where 
space appears to satisfy Euclid's postulates to excellent approximation.

We therefore have to assume all the varied contributions to the cosmological 
term cancel to very high precision for some mysterious reason. We will 
assume this from now on to make contact with the observed universe. This 
unnatural feature is called the Cosmological Constant Problem \cite{ccp}. 

In the modulus stabilization example of the last section, the value of 
$V_{\rm eff}$
 at its minimum represents (in general, a contribution to) the infrared 4D 
cosmological 
constant. As pointed out at the end of that section, 
in the orbifolded version one can also add an $R$-independent constant to 
$V_{\rm eff}$, originating from a boundary-localized constant potential. 
By fine-tuning the value of this constant we can ensure that the minimal
$V_{\rm eff}$ vanishes, that is, the 4D effective cosmological constant 
is zero (or very small). The ugliness
of having to do this tuning reflects the 
unresolved cosmological constant problem.

Just like 4D General Relativity, 5D General Relativity can also 
have a cosmological term, 
\be
S=\intc d^5X \sqrt{G}\left\{\frac{\RC}{G_N^{5D}}-\La \right\}.
\ee
Now what we have learned in gauge theory is generally true: 4D effective 
couplings derive from 5D couplings, but they are not the same thing. 
In particular, our tuning of the 4D cosmological term, $\lambda$, to be 
very nearly zero does not indicate that $\Lambda$ is near zero (in fact 
we will see this explicitly below). The most natural thing is to therefore 
assume that $\Lambda \neq 0$ and see what happens. 

In our 5D cylinder example, let us treat $\Lambda$ as a small perturbation, 
where at zeroth order we have a cylindrical solution to Einstein's 
equations, 
\be
\mean{\,ds^{\,2}} =
\eta_{\mu\nu}dx^\mu dx^\nu -R_\text{phys}^2 \,d\phi^2 \; .
\ee
Therefore the cosmological term in the 5D action becomes
\bea
S \ni \intc d^4x\!\!\int_0^{2\pi} \bbb\!d\phi \,R (-\La)
=\intc d^4x (-2\pi R\La) \; ,
\eea
which looks precisely like one of our old contributions to the radion ($R$) 
effective potential. In fact we saw that there were divergent 
renormalizations of this $\Lambda$ from 5D matter loops, again 
suggesting that the renormalized $\Lambda$ should naturally be significant. 

Once we decide that  $\Lambda$ is not very small, we really should consider 
its effects in determining the 5D geometry. It cannot be treated 
perturbatively, we must resolve Einstein's equations in the presence of the 
5D cosmological constant.
The same is true for any boundary-localized potential terms coupled to 
5D gravity.

\section{Warped Compactification}

We will study the simplest model where we include the higher dimensional 
cosmological term, known as the Randall-Sundrum I (RS1) model \cite{rs1} 
\cite{rs2}. It is also a 
good prototype of more complex constructions.

Rather than continue with the cylindrical spacetime we return to the 
orbifolded variant. The 5D gravitational action of the ``bulk'' of the 
spacetime is still given by 
\be
S_\text{bulk} =\intc d^4x\int_{-\pi}^{+\pi} \bbb d\phi\,
\sqrt{G}\left\{\frac{\RC}{G_N^{5D}}-\La \right\}.
\ee
We can assign the orbifold parities as $+1$ for $G_{\mu \nu}$ and  
$G_{55}$ and 
$-1$ for $G_{\mu 5}$, which is respected by the action. In almost 
axial gauge then, $G_{\mu 5}^{(0)}(x) = 0$ by the parity. Therefore in the 
orbifolded set-up we automatically have no KK massless gauge boson 
emerging in the 4D effective theory. Of course there can be even more 
dimensions in non-minimal set-ups from which such gauge bosons might emerge. 
Here, when we need gauge bosons we will just add them at the 5D level. 
The radion does remain after orbifolding since $G_{55}$ is parity even.

The orbifolded set-up also has ``branes'' or boundaries of 5D spacetime, 
namely the 4D subspaces at $\phi = 0, \pi$. In general there can be 
4D actions localized to these branes for the 5D fields. In fact such 
actions must be there since they are renormalized by 5D loops \cite{hailu}. 
The leading 
terms in these brane actions are ``tensions'', which look like 
localized cosmological terms, 
\be
S_{\text{brane}(i)}= -\intc d^4x\sqrt{-g^{(i)}}T^{(i)} \; ,
\ee
where 
\bea
&&g_{\mu\nu}^{(1)}(x)=G_{\mu\nu}(x,\phi=0) \nn\\
&&g_{\mu\nu}^{(2)}(x)=G_{\mu\nu}(x,\phi=\pi) \; ,
\eea
and the $T^{(i)}$ are constant tensions.
The induced 4D metrics define distances along the branes, for example, 
\be
ds_{(1)}^{\,2} =G_{\mu\nu}(x,\phi=0)dx^\mu dx^\nu \; ,
\ee
since $d \phi = 0$ along the brane.

Since we are looking for solutions to Einstein's equations that 
might fit the vacuum of the 
real world, let us try the ansatz that the 5D metric should 
respect at least 4D Poincare invariance, 
\be
ds^{\,2} = e^{-2\si(\phi)}\eta_{\mu\nu}dx^\mu dx^\nu
-R^2 d\phi^2 \; .
\ee
Here, $\eta_{\mu \nu}$ is the 4D Minkowski metric, and 
we have chosen the extra-dimen\-sional coordinate to be 
proportional to proper distance. The prefactor to $\eta_{\mu \nu}$ is 
written as an exponential as a convenient convention and is called the
``warp factor''. Its potential $\phi$-dependence means that the 
higher-dimensional geometry cannot be defined as a product geometry of 
4D Minkowski space and some purely extra-dimensional geometry, but 
rather all the dimensions are entangled. Plugging this ansatz into 
the equations of motion following from our bulk plus brane actions,
one finds
\bea
6\,{\si^\pr}^{\,2} &=& -\frac{\La}{4M_5^3} \equiv 6k^2 \nn\\
3\,\si^{\pr\pr} &=&
\frac{T^{(1)}}{4M_5^3}\,\da(R\phi) +
\frac{T^{(2)}}{4M_5^3}\,\da(R(\phi-\pi)) \; ,
\eea
where we define a 5D Planck scale,
\be
M_5^3 \equiv \frac{1}{2\, G_N^{5D}} \; .
\ee
The only consistent solution to these equations, satisfying periodicity 
in $\phi$ and the orbifold parity is illustrated in Fig. 16. 
\bfi[tb]
\pg{a}{$-\pi$}
\pg{b}{$\pi$}
\pg{c}{$\phi$}
\pg{d}{$\sigma$}
\begin{center}
\includegraphics{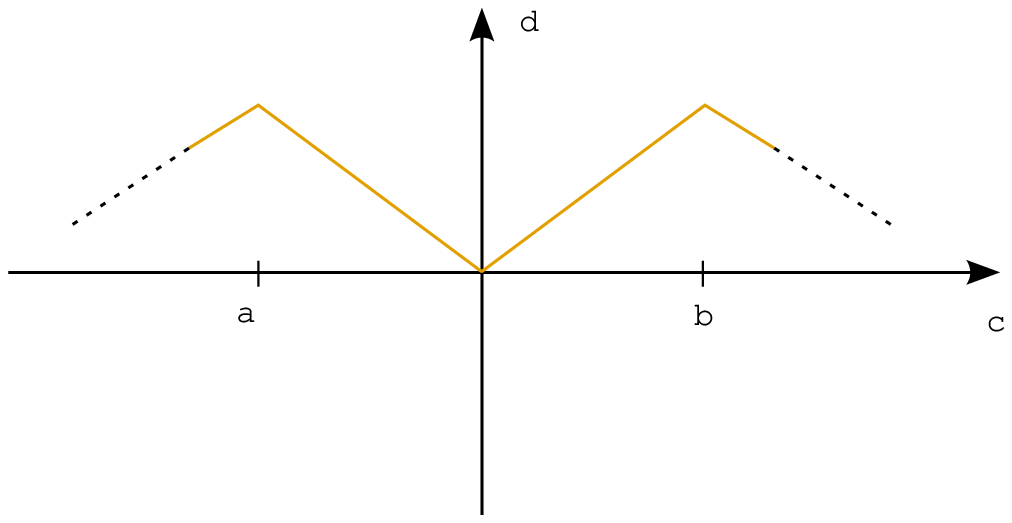}
\end{center}
\caption{Warp factor solution satisfying orbifold boundary condition}
\efi
But even this 
solution only exists if the kinks have the right size to reproduce the 
$\delta$-functions in the equations of motion. This requires the 
relationships between brane tensions and bulk cosmological constant given by
\be
T^{(1)}=-T^{(2)}=24k M_5^3 \; .
\ee
Thus the vacuum metric solution is given by 
\bea
ds^{\,2} &=& e^{-2k R\phi}\eta_{\mu\nu}dx^\mu dx^\nu -R^2 d\phi^2, \quad
0\leq\phi\leq\pi \nn\\[2mm]
&{\ba[t]{c} =\\[-2mm] \sst y \,\equiv\, R\phi \ea}&
e^{-2k y}\eta_{\mu\nu}dx^\mu dx^\nu-dy^2, \quad
0\leq y\leq \pi R \; .
\eea
This geometry is a slice of $AdS_5$, 5D anti-de Sitter spacetime, 
which is the maximally symmetric 5D spacetime of negative curvature (just 
as a sphere is the maximally symmetric space (not spacetime) 
 of positive curvature). 

The forms of the 
massless 4D modes, the 4D graviton and radion, are easy to guess: 
\be
ds_{(0)}^{\,2} =e^{-2k R(x)\phi}g_{\mu\nu}^{(0)}(x)dx^\mu dx^\nu
-R(x)^2d\phi^2, \quad 0\leq\phi\leq\pi \; .
\ee
That is we promote the radius, which thusfar has been an integration 
constant of our vacuum solution, into an $x$-dependent field, and 
$\eta_{\mu \nu}$ into a dynamical $4$-metric, $g^{(0)}_{\mu \nu}(x)$. 
There is a 
slick proof that these fluctuations indeed do correspond to the 4D massless 
modes, that indeed they have no potential at all classically. Plugging the 
above ansatz into our full action must give a vanishing
 effective 4D cosmological term since we found a 4D Poincare invariant 
solution. Now a 4D potential for $g^{(0)}_{\mu \nu}$ and $R$ can be computed 
for $x$-independent fields. A constant linear coordinate transformation 
can take any non-singular symmetric matrix, $g^{(0)}_{\mu \nu}$, to the 
form $\eta_{\mu \nu}$. Therefore by residual 4D general coordinate 
invariance of the action, plugging $x$-independent $g^{(0)}_{\mu \nu}, R$ 
must also give zero. Since the effective lagrangian  is the negative of the  
effective potential for constant fields, the effective potential for 
$g^{(0)}_{\mu \nu}, R$ clearly vanishes. So we have identified the 
zero-modes and the radion is a modulus again. 

Radius stabilization can be readily achieved much as in the unwarped case 
we studied earlier for the cylindrical spacetime, by adding suitable 5D 
matter. The simplest such implementation is the classical 
Goldberger-Wise mechanism \cite{gw} 
using a 5D scalar, naturally allowing a moderately 
large ``radius'',
\be
k\mean{R} \sim \mathcal{O}(10) \; ,
\ee
where $k$ ($\Lambda$) is taken not too much smaller than the 5D effective 
field theory cutoff. We could also proceed with stabilization by 
quantum corrections from massive matter as we studied earlier in the 
unwarped case. We will not pursue these in detail here, just assume some 
such stabilization, the warping has mild impact on the stabilization.

Ex. Work through Goldberger-Wise stabilization \cite{gw}.

After radion stabilization, and neglecting the small backreaction of the 
stabilizing physics on the vacuum metric solution, we have only 
the 4D metric zero mode, 
\be
ds_{(0)}^{\,2} =e^{-2k \mean{R}\phi}g_{\mu\nu}^{(0)}(x)dx^\mu dx^\nu
-\mean{R}^2d\phi^2, \quad 0\leq\phi\leq\pi \; .
\ee

The 4D low-energy effective action for the 4D metric zero mode is 
obtained by plugging the above ansatz into the fundamental action. The 
5D curvature term contains 2-derivative terms, 
which by 4D Poincare invariance of the vacuum solution must be either 
two $x$-derivative acting on the zero mode or two $\phi$ derivatives 
acting on the warp factor. From the 4D point of view all $\phi$ 
derivatives just contribute to potential terms, which we saw above 
all cancelled for the 4D graviton zero-mode. So we focus on pairs of 
$x$-derivatives. Plugging in the zero-mode ansatz, 
a typical term in the fundamental action has the schematic form,
\bea
G_N^{5D}S &\ni& 
\intc d^4x \!\!\int_0^\pi \!\!d\phi \,\sqrt{G}
G^{xx}G^{xx}G^{xx}\p_x G_{xx}\p_x G_{xx} 
\\ &=&
\intc d^4x \!\!\int_0^\pi \!\!d\phi \,R e^{-2k R\phi}
\sqrt{-g^{(0)}}g^{(0)xx}g^{(0)xx}g^{(0)xx}
\p_x g^{(0)}_{xx}\p_x g^{(0)}_{xx} \; , \nn
\eea
which allows us to count powers of the warp factor easily. 
The residual 4D general coordinate invariance then gives a unique form 
for the 4D effective action of the graviton zero-mode, 
\bea
S_{4D \,\text{eff}} &=& \frac{1}{G_N^{5D}}
\intc d^4x \left(\int_0^\pi \!\! d\phi\, R e^{-2k R\phi}\right)
\sqrt{-g^{(0)}}\,\RC_{(4D)}[g^{(0)}] 
\nn\\ &=&
\frac{1}{2k G_N^{5D}}
\left(1-e^{-2k\pi R}\right)
\intc d^4x\sqrt{-g^{(0)}}\,\RC_{(4D)}[g^{(0)}] \; .
\eea
That is, at low energies we obtain 4D General Relativity. 

We easily deduce the 4D effective Newton's constant, 
\be
16\pi G_N^{4D\,\text{eff}} =\frac{2k \,G_N^{5D}}{1-e^{-2k\pi R}} \; .
\ee
Remarkably, even when we take the decompactified limit, 
$R \rightarrow \infty$, this effective coupling stays finite, 
\be
M_{4 Pl}^{\,2} =\frac{M_5^3}{2k}\left(1-e^{-2k\pi R}\right)
{\ba[t]{c} \longrightarrow \\[-2mm] \sst R\rightarrow\infty \ea}
\frac{M_5^3}{2k} \; .
\ee
This is a result of the fact that the zero-mode is localized in the 
vicinity of the brane at $\phi = 0$. This localization mechanism for gravity 
is sometimes known as the Randall-Sundrum II (RS2) mechanism. 
Because of gravity localization, the brane at $\phi = 0$ is usually called 
the ``Planck brane''.

All these classical derivations are considered to take place as the 
leading approximation of non-renormalizable quantum effective field theory. 
The control parameter is in general given by
\be
\frac{E^3}{M_5^3}\ll 1
\ee
in particular, 
\be
\frac{\La^{3/5}}{M_5^3} \sim
\frac{(M_5^3 k^2)^{3/5}}{M_5^3} \ll 1
\quad\Longleftrightarrow\quad
\frac{k}{M_5} \ll 1 \; .
\ee
In practice we usually take all the scales within an order of magnitude or 
two of the 4D Planck scale, 
\be
k \;\lab\; M_5 \;\lab\; M_{4\,Pl} \;\sim\; 10^{18}\text{GeV} \; .
\ee

\section{Warped Hierarchy}

Consider the set-up of Fig. 17, 
\bfi[tb]
\pg{a}{4D}
\pg{b}{$y=0$}
\pg{c}{$y=\pi R$}
\pg{d}{\bt{c} Warp factor\\ (profile of \\ 4D graviton) \et}
\pg{e}{$H(x)$}
\begin{center}
\includegraphics{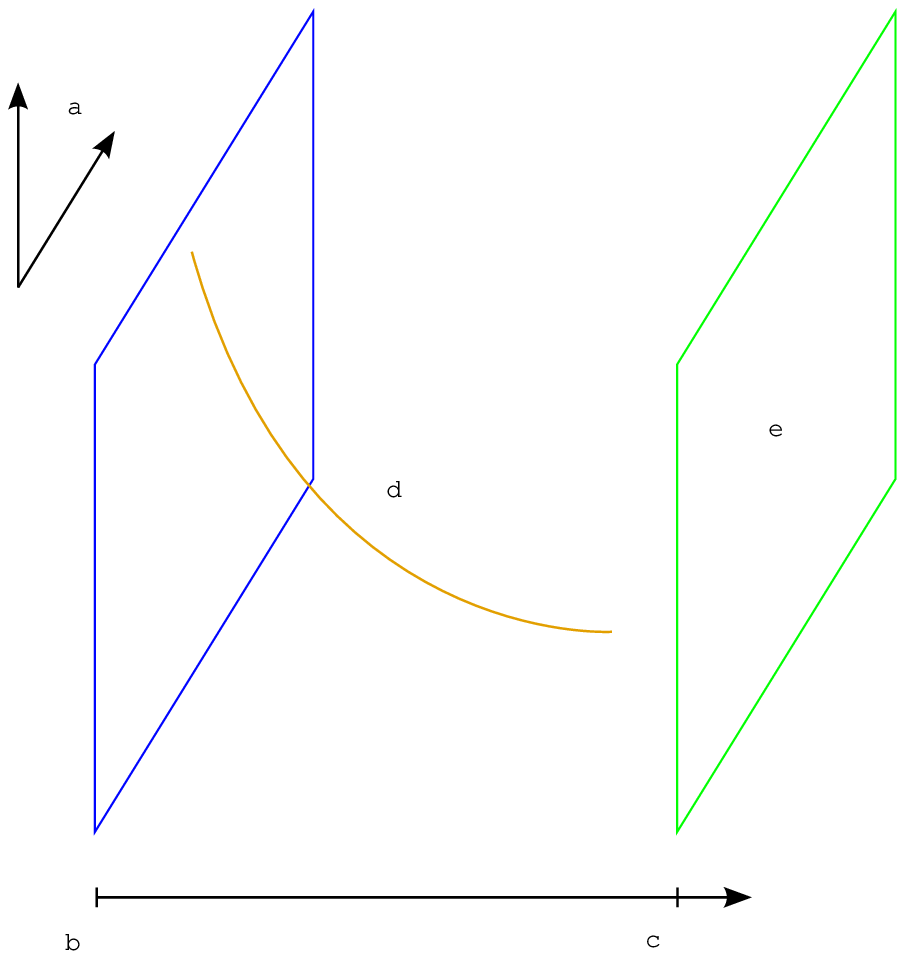}
\end{center}
\caption{Localization of 4D graviton and Higgs}
\efi
where we have a warped compactification with 
localized 4D gravity, and now we add a 4D Higgs field by hand (not from 
$A_5$'s for simplicity), 
$\delta$-function localized to the opposite brane ($\phi =  \pi$), with 4D 
brane-localized action, 
\bea
S_\text{Higgs} = \intc d^4x\sqrt{-g_\text{ind}} \left\{
g^{\mu\nu}_\text{ind}\p_\mu H^\dagger \p_\nu H
-\la (|H|^2-v_0^2)^2 \right\},
\eea
where 
\bea
g_{\mu\nu}^\text{ind}(x) &=&
G_{\mu\nu}(x,\phi=\pi) 
\nn\\[2mm] &=&
e^{-2k\pi R}g_{\mu\nu}^{(0)}(x) \; ,
\eea
gives the induced 4D geometry of the brane, and  where the second line 
gives the low-energy approximation where only 
the gravity zero-mode can propagate. In this approximation the Higgs action 
becomes
\be
S_\text{H} =\intc d^4x \sqrt{-g^{(0)}} \left\{
e^{-2k\pi R}g_{(0)}^{\mu\nu}\p_\mu H^\dagger \p_\nu H
-e^{-4k\pi R}\la (|H|^2-v_0^2)^2 \right\},
\ee
where the warp factor appears like a conventional constant wavefunction 
renormalization of the Higgs. Canonical normalization of the Higgs is 
achie\-ved by the field redefinition, 
\be
e^{-k\pi R}H \rightarrow H \; ,
\ee
giving a canonical action, 
\be
S_\text{H} =\intc d^4x \sqrt{-g^{(0)}} \left\{
g_{(0)}^{\mu\nu}\p_\mu H^\dagger \p_\nu H
-\la \left(|H|^2-
\underbrace{e^{-2k\pi R}v_0^2}
\right)^2 \right\}.
\ee
The miracle is that the ``bare'' weak scale parameter for the Higgs, 
$v_0$, receives a large warp factor renormalization reducing it to the 
physical weak scale Higgs VEV,
\be
v = e^{-k\pi R}v_0 \; .
\ee
Even if $v_0$ is very large, Planckian in size, for $kR \sim {\mathcal O}(10)$ 
one can easily accomodate the physical weak scale, $v \sim 250$ GeV. 
Because of this large warping down of UV scales, this brane is sometimes 
called the IR brane, or in the present context, the ``TeV brane''.

How does this magic trick work? Consider again the vacuum solution for the 
5D metric, 
\be
ds_\text{vacuum}^{\,2} = e^{-2k y}\eta_{\mu\nu}dx^\mu dx^\nu
- dy^2 \; .
\ee
In any extra-dimensional locale, $y \sim y_0$, we can approximate this 
metric, 
\bea
ds_\text{vac}^{\,2} &\sim& 
e^{-2k y}\eta_{\mu\nu}dx^\mu dx^\nu - dy^2
\nn\\ &=&
\eta_{\mu\nu}d\hat{x}^\mu d\hat{x}^\nu - dy^2 \; ,
\eea
which looks just like a piece of 5D Minkowski spacetime provided 
 we define
\be
\hat{x} \equiv e^{-k y_0}x, \quad
\wh{m}_{4D} \equiv e^{k y_0} m_{4D} \; .
\ee
Without warping the translation from 5D to 4D masses is straightforward, 
so for physics localized in the vicinity of $y_0$, we have the 
translation, 
\be
\wh{m}_{4D} \sim m_{5D} 
\quad\Longrightarrow\quad
m_{4D} \equiv e^{-k y_0} m_{5D} \; .
\ee
For the case of the Higgs, 
\be
y_0 =\pi R \; ,
\ee
so
\be
v = e^{-k\pi R} v_0 \; .
\ee

Warped hierarchies are radiatively stable, essentially because even 
(general coordinate invariant) cutoff scales get warped down near the IR 
brane, as illustrated in Fig. 18.
\bfi[tb]
\pg{a}{\bt{c} Planck \\ brane\et}
\pg{b}{IR brane}
\pg{c}{\bt{c} 
Regulator fields\\[2pt] for 5D GR \\[2pt]
mass $\La_5^\text{grav}\sim M_\text{Pl}$
\\[5pt]
$\La_4^\text{grav} \sim e^{-k y}M_\text{Pl}$ 
\et}
\pg{d}{\bt{c} 
Pauli-Villars\\[2pt] for Higgs \\[2pt]
mass $\La_5^\text{H} \sim M_\text{Pl}$
\\ [5pt]
$\La_4^\text{H} \sim e^{-k\pi R}M_\text{Pl}$ 
\et}
\hspace{20pt}\mbox{
\includegraphics{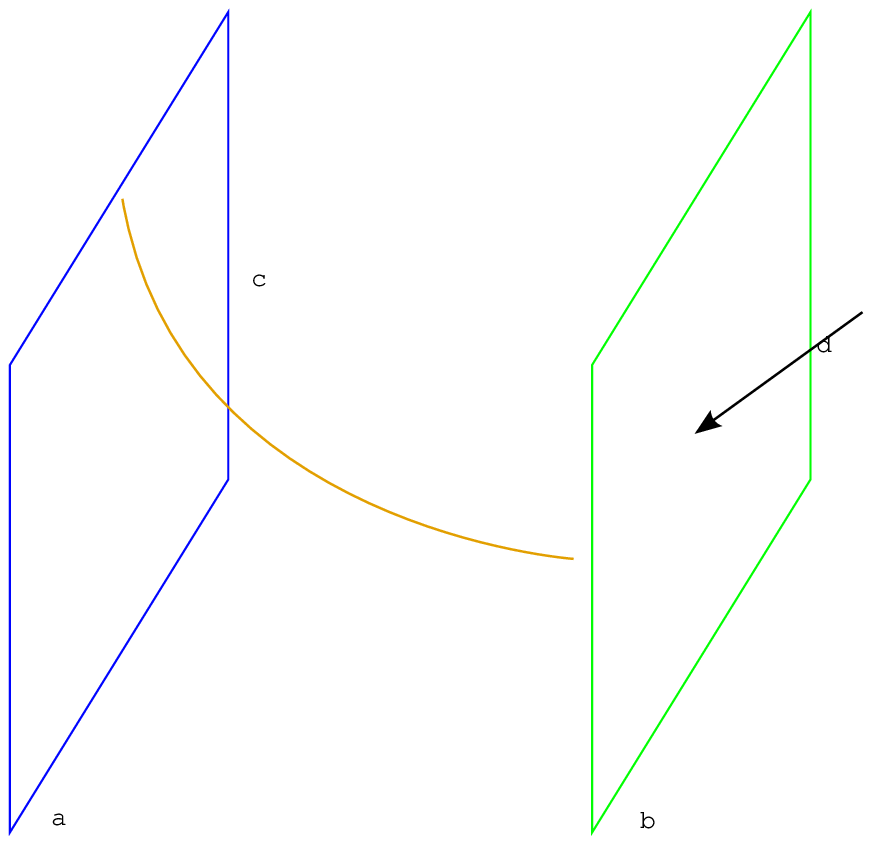}
}
\caption{Profiles of 5D regulator fields for gravity
and 4D regulator for Higgs}
\efi

\section{KK Gravitons at Colliders}

Let us begin by removing the IR brane to infinity, $R \rightarrow \infty$, 
and expand the general 5D metric (in almost axial gauge) about the 
vacuum solution, 
\be
ds^{\,2}=\left[
e^{-2k|y|}\eta_{\mu\nu}+e^{-k|y|/2}h_{\mu\nu}(x,y)\right]
dx^\mu dx^\nu - dy^2 \; .
\ee
The parametrization of fluctuations with the extra exponential is purely 
for later convenience.
It is also convenient to switch to an extra-dimensional coordinate, 
\be
z \equiv \sgn(y)\frac{e^{k|y|}-1}{k} \; .
\ee
Plugging the general 5D metric into the fundamental action, and expanding to 
quadratic order in  small fluctuations about the vacuum gives, 
\be
G_N^{5D} S = 
\intc d^4x \bb\int\limits_{-\infty}^{+\infty}\bb dz
\left\{
\frac{1}{2}h_{\mu\nu}\Box^{\,\mu\nu\rho\si}h_{\rho\si}
-\frac{1}{2}h_{\mu\nu}\eta^{\,\mu\nu\rho\si}
H_\text{QM}h_{\rho\si}
\right\},
\ee
where the first term contains an operator made of two $x$-derivatives
 while the second term contains an operator made from $y$-derivatives 
and non-derivative terms, specified by 
\be
H_\text{QM} = -\frac{1}{2}\p_z^{\,2} + V_\text{QM}(z), \quad
V_\text{QM}(z) \equiv
+\frac{15k^2}{8(k|z|+1)^2}
-\frac{3k}{2}\da(z) \; .
\ee
The unspecified index structures are exactly such that, if 
$H_{QM}$ is only a constant and if the fields are purely 4D fields, 
 and if we get rid of the $z$ integral, 
this action would just be that of a free {\it massive} 4D spin-2 
field, with mass-squared given by the constant  $H_{QM}$. Of course, 
 $H_{QM}$ is a hermitian operator in the extra dimension, so we should really 
try to diagonalize the operator. The eigenvalues will then
 be the 4D mass-squareds of the tower of gravitational KK modes.
Note that this eigenvalue problem is analogous 
to finding the energy eigenvalues of a 1D non-relativistic  quantum 
mechanics problem with ``unit'' mass and a ``volcano potential'', 
\be
V_\text{QM}(z) \equiv
+\frac{15k^2}{8(k|z|+1)^2}
-\frac{3k}{2}\da(z) \; ,
\ee
as illustrated in Fig. 19. 
\bfi[tb]
\pg{a}{$z$}
\pg{b}{$V_\text{QM}$}
\begin{center}
\includegraphics{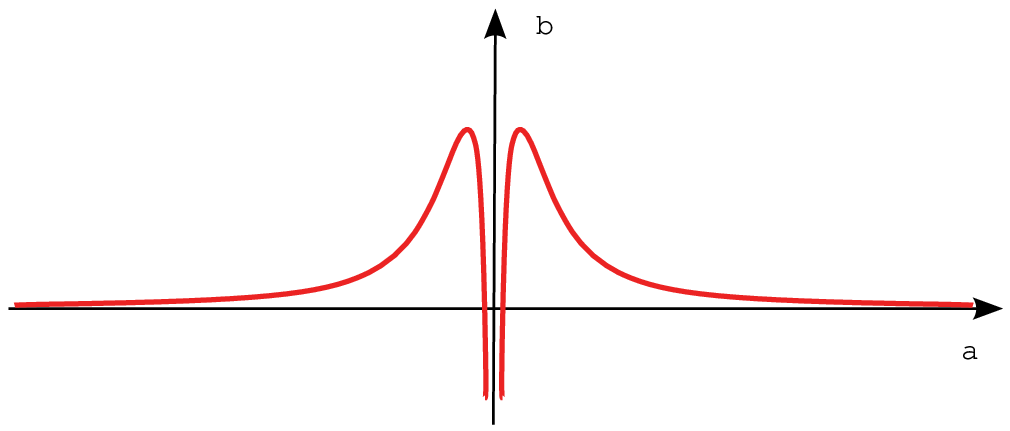}
\end{center}
\caption{``Volcano'' potential of analog quantum mechanics
problem describing KK graviton fluctuations}
\efi

By inspection the analog time-independent Schrodinger equation should 
have precisely one bound state (which of course must therefore be 
noneother than our localized zero-mode) and a continuum of waves 
(asymptoting to sinusoidal waves far away from the volcano) with 
positive ``$E_{QM}$'' $>0$. Now let us put back in the fact that 
$R < \infty$. In the $z$ coordinates, 
\be
|z_\text{max}| \sim {e^{k\pi R}}/{k} \; .
\ee
While this does little to the bound state mode, it will quantize the 
previously continuum modes, with 
\be
\text{``}E_\text{QM}\text{''} \sim (n/z_\text{max})^2 \; .
\ee
From this we deduce that the zero mode graviton is accompanied by KK 
spin-2 excitations with masses given by
\be
m_\text{KK}^2 \sim \left(n k e^{-k\pi R} \right)^2 \; .
\ee
Even though all the input scales of the fundamental 5D set-up are 
Planckian, given that the warp factor renormalized the weak scale to its 
observed size, we get an estimate for KK gravitons, 
\be
m_\text{KK} \lab n M_{Pl}e^{-k\pi R} \sim \text{TeV !}
\ee
That is, these states are accessible to TeV scale colliders.

Of course it is not good enough to be kinematically accessible, to be seen 
a new particle must have an appreciable coupling to ordinary matter. If 
ordinary matter couples to KK gravitons with the same strength as the usual 
graviton, these states would be completely invisible as a practical matter.
Fortunately this is not the case.
To be concrete let us assume that all matter, including the Higgs is 
localized on the TeV brane. They can only couple to the graviton KK modes via 
their extra-dimensional profiles evaluated on the TeV brane. This determines 
the relative strength of coupling of different gravity KK modes to brane 
localized matter since the 
fundamental coupling constant, the 5D Newton constant, is the same for all 
the KK modes. The relative strength is easy to estimate. 
A typical low-lying KK gravity excitation will have roughly a 
plane-wave wave-function (for the analog non-relativistic quantum mechanics 
problem) in most of the $z$ space, so that its normalized value 
on the IR brane is roughly of order $1/\sqrt{z_{max}} 
\sim \sqrt{k} e^{-k \pi R/2}$. On the other hand, 
the bound state 4D massless graviton mode has the normalized 
analog wavefunction evaluated
on the IR brane given by $\sim \sqrt{k} e^{- 3 k \pi R/2}$. 

Ex. Check this last fact carefully accounting for the factor of $3/2$ in the 
exponent arising from the precise definition of the analog problem in terms 
of the 5D metric ansatz.

Thus the amplitude for a KK excited graviton to couple to brane matter is 
$\sim e^{k \pi R}$ larger than the amplitude for the massless graviton. Now 
the massless graviton of course couples with 4D Planck-suppressed 
$\sim 1/(10^{18} {\rm GeV})$ strength which is why such couplings are 
invisible at TeV scale colliders. But since we are taking $e^{k \pi R} 
\sim M_{Pl}/$TeV, KK gravitons will couple to matter with only $1/$TeV 
suppression, that is order one amplitudes at TeV colliders \cite{hooman}! 
In this 
scenario these exotic 
spin-2 states should be visible at upcoming colliders such as the LHC.
Similar mass splitting, wavefunctions and coupling strengths to IR brane 
matter hold for KK excitations of any 
field (with any spin) one considers in the 5D bulk.

\section{Warped (Fermionic) Bulk Matter}

A useful and pioneering reference for this section and the next is 
Ref. \cite{tony}.
Let us consider 5D fermions in the warped context subject to the orbifold 
boundary conditions,
\bea
P(\PsiL) &=& +\PsiL \nn\\ 
P(\PsiR) &=& -\PsiR \; .
\eea
Using the proper-distance $y$ coordinate ranging from $- \pi R$ to $\pi R$, 
we consider a bulk action of the form
\be
S_\Psi =\intc d^4x\bb\int\limits_{-\pi R}^{+\pi R}\bb dy\,
e^{-4k|y|}\Psibar\left(
i\,\Ga^M D_M -m \;\sgn(y) \right)\Psi \; .
\ee
In general, the covariant derivative for fermions cannot be written directly 
in terms of the metric alone, but requires a compatible ``spin-connnection''.
We will just consider the RS1 vacuuum metric for which the compatible 
covariant derivatives are given by 
\bea
D_\mu &=& \p_\mu -\frac{1}{2}\Ga_5\Ga_\mu k\;\sgn(y) \nn\\
D_5 &=& \p_y \; .
\eea
The exponential is just the measure factor arising from the root of the 
metric determinant as usual, but it appears that the mass term has 
explicit $y$ dependence in order to be compatible with the orbifold parity 
symmetry. This is not really cheating, such a mass term can be thought of as 
arising from a Yukawa coupling to a orbifold-parity-odd 5D scalar field, 
whose fluctuations are very massive but whose VEV is non-zero and proportional 
to sgn$(y)$. 
The curved space $\Gamma^N$ are 
given in terms of the usual flat space Dirac matrices by 
$\Gamma_\mu = e^{-k|y|}\ga_\mu, ~\Gamma_5 =-i \gamma_5,
~\Gamma^\mu = e^{+k|y|}\ga^\mu, ~\Gamma^5 = i \gamma_5$.

Decomposing the action in 4D notation, 
and making the convenient field redefinition, 
$\Psi \equiv e^{+ 3/2 k |y|} \hat{\Psi}$,
\be
S_\Psi =\intc d^4x\bb\intc dy\
\hat\Psibar \left\{
i\!\xd +e^{-k |y|}\left[\frac{k}{2}\,\sgn(y)\ga_5
-\ga_5\p_y -m\,\sgn(y)\right] \right\}
\hat\Psi \; .
\ee
Despite the 5D mass parameter, which we will from now on 
dimensionlessly parametrize as
\be
c \equiv \dst\frac{m}{k} \; ,
\ee
there are 4D massless chiral fermion zero-modes. We clearly get a zero-mode 
from the equation of motion if the mode is annihilated by the 4D Dirac 
operator, so that
\be
\left[k\left(\frac{\ga_5}{2} -c\right)\sgn(y)-\ga_5\p_y\right]
\hat\Psi =0 \; .
\ee
We see two possible chiral solutions to this equation. The left-handed one, 
\be
\hat\Psi_\mathrm{L}{}(x,y) 
=\hat\Psi_\mathrm{L}{}^{\bb(0)}(x) \,e^{(\frac{1}{2}-c)k|y|} \; ,
\ee
satisfies the orbifold parity condition and is therefore physical. The 
right-handed one, 
\be
\hat\Psi_\mathrm{R}{}(x,y) 
=\hat\Psi_\mathrm{R}{}^{\bb(0)}(x) \,e^{(\frac{1}{2}+c)k|y|} \; ,
\ee
does not satisfy being parity-odd, and therefore is inadmissible. The parity 
therefore gives us a chiral 4D massless left-handed zero-mode. 

While the 5D mass parameter, $c$, does not affect the existence of 4D 
chiral modes, they clearly influence their profile in the extra dimension, 
as illustrated in Fig. 20. 
\bfi[tb]
\pg{a}{\bt{c} Planck \\ brane\et}
\pg{b}{IR brane}
\pg{c}{$c>1/2$}
\pg{d}{$c<1/2$}
\pg{e}{\bt{c} H\\Yukawa\\couplings \et}
\hspace{50pt}\mbox{
\includegraphics{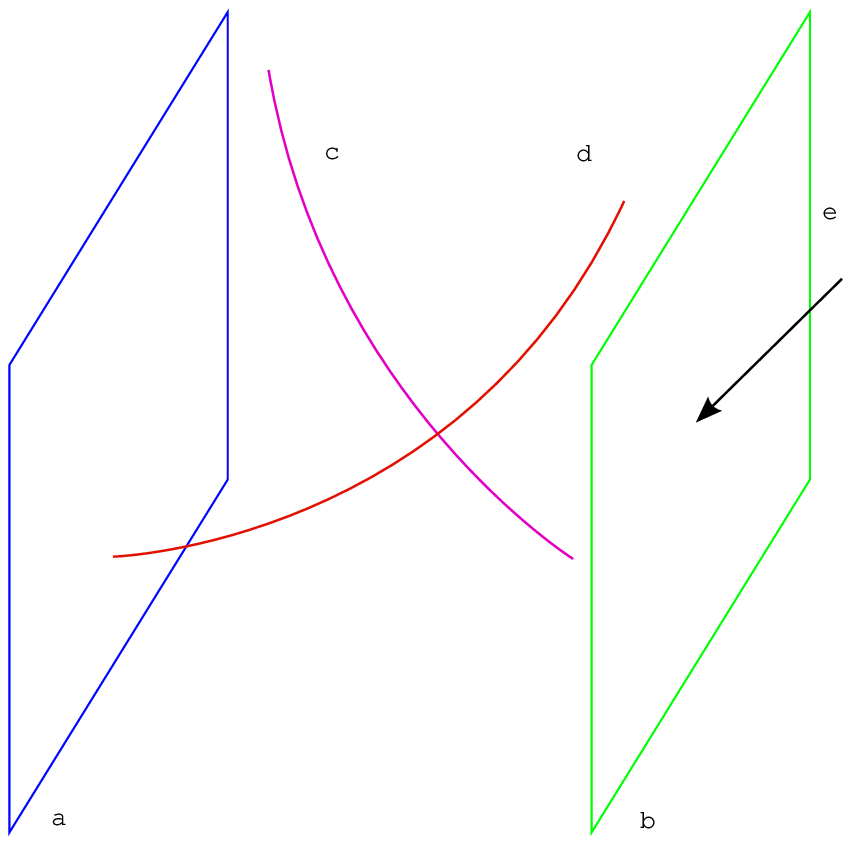}
}
\caption{Fermion zero-mode profiles for different
5D fermion masses }
\efi
If the Higgs is considered to be localized on the 
IR brane still, 4D Yukawa couplings with two species of chiral zero modes coming 
from bulk fermions with mass parameters $c_a, c_b$ will be given by 
\be
{\ba[c]{c} 4D \;\text{Yukawa} \\ \text{couplings} \ea} \sim
\,e^{\tst(\frac{1}{2}-c_a)k\pi R} 
\;e^{\tst(\frac{1}{2}-c_b)k\pi R}
\times {\ba[c]{c} 5D \;\text{Yukawa} \\ \text{couplings} \ea}.
\ee
Thus even without large hierarchies at the 5D level, hierarchical 
effective Yukawa couplings are naturally generated. In the real world, 
we can identify light fermions as chiral modes arising from bulk fermions with 
$c > 1/2$, and heavy fermions with chiral modes arising from bulk fermions 
with $c< 1/2$. Therefore light fermion profiles are suppressed at the IR brane. 
This suppresses their wave-function overlap with low-lying 
KK excitations of all bulk fields, thereby suppressing a host of dangerous 
KK-mediated effects. This is the central part of an automatic GIM mechanism 
suppressing flavor-changing neutral currents. On the other hand one can 
predict that the heavy top quark in this scenario should display significant 
non-standard corrections to its couplings.

\section{Warped Bulk YM}

Of course if fermionic matter lives in the 5D bulk so must the gauge fields. 
The 5D YM action in background curved space is given by 
\be
S =-\frac{1}{4}\intc d^4x \bb\int\limits_{-\pi}^{+\pi}\bb d\phi\,
\sqrt{G}G^{MN}G^{KL}F_{MK}F_{NL} \; .
\ee
We will consider the RS warped vacuum background.

We begin by considering the orbifold boundary conditions given by
\bea
P(A_\mu) &=& +A_\mu \nn\\
P(A_5) &=& -A_5 \; .
\eea
In almost axial gauge, the single expected $A_5$ mode is eliminated by the 
orbifold conditions, but $A_{\mu}$ does give rise to a zero-mode.

Ex. Show that
\be
A_\mu(x,\phi)=A_\mu^{(0)}(x)
\ee
is the zero-mode 4D gauge boson and that if coupled to matter (or if 
non-abelian), the effective 4D gauge coupling is given by 
\be
g_4=\frac{g_5}{\sqrt{2\pi R}} \; .
\ee

Note the form of the zero-mode and the 4D coupling look exactly as they did 
without warping. Gauge theory is very different from gravity in this regard.
In particular the warped explanation for ``why gravity is weak'' (the 
zero-mode graviton of course) does not simultaneously greatly weaken the 
strength of gauge forces, which is obviously a good thing. 

The other interesting example to consider is given by orbifold conditions,
\bea
P(A_\mu) &=& -A_\mu \nn\\
P(A_5) &=& +A_5 \; ,
\eea
so that now there is a 4D scalar zero-mode but not a vector zero-mode. 
(Of course, as illustrated earlier in flat spacetime, for non-abelian gauge 
group different gauge fields corresponding to different generators can 
be treated with different orbifold boundary conditions. We will not consider 
that level of complexity here.)
To isolate the scalar zero-mode in the warped background we first 
look at all terms in the action containing $A_5$,
\bea
S &\ni& 
\frac{1}{2}\intc d^4x \bb\intc d\phi\,
R \,e^{-4k|\phi| R} e^{2k|\phi| R} \left(
\p_5 A_\mu -\p_\mu A_5 \right)^2 
\nn\\[2mm] &\ni&
\frac{1}{2}\intc d^4x \bb\intc d\phi\,
R \,e^{-2k|\phi|R} \left[
(\p_\mu A_5)^2 -2\,\p_5 A_\mu \p^\mu A_5 \right].
\eea
The usual 
almost axial gauge does not disentangle the vector-scalar mixing in warped 
spacetime and is 
therefore not convenient. But if we try 
\be
A_5(x,\phi) =A_5^{(0)}(x) \,e^{+2k|\phi| R} \; ,
\ee
the mixing term vanishes (as you can check by doing an integration by 
parts with respect to the $\phi$ derivative). The 4D effective action for 
the zero-mode scalar below the mass of the lightest KK excitations (
$\sim k e^{-k \pi R}$) is then given by 
\bea
S_{\text{eff}\,4D} &=&
\frac{1}{2}
\intc d^4x \bb\int\limits_{-\pi}^{+\pi}\bb d\phi\,
R \,e^{+2k|\phi|R} \left(\p_\mu A_5^{(0)} \right)^2
\nn\\[2mm] &=&
\intc d^4x \,\frac{e^{2k\pi R}-1}{2k}
\left(\p_\mu A_5^{(0)} \right)^2 \; .
\eea

As the zero-mode ansatz shows such massless scalars are localized (but not 
$\delta$-function localized) near the IR brane and therefore makes a good 
Higgs candidate \cite{comphiggs}! 

One can ask what the advantage is of having the Higgs realized as the 
fifth component of a gauge field as opposed to as a fundamental 
IR-brane-localized scalar. The answer is that in the former case 
radiative corrections to the scalar mass-squared are effectively cut off  
at $\sim m_\text{KK} \sim k e^{-k \pi R}$ while in the latter case the 
cutoff is the ``warped down'' fundamental cutoff 
$\Lambda_\text{UV} e^{- k \pi R}$, which is parametrically larger. 
This is a generalization of the flat spacetime result we 
derived in detail earlier where the cutoff on the scalar mass was 
effectively $m_\text{KK} \sim 1/R$.
One can repeat the kind of calculations we did at the 
beginning of these lectures which carve out an $A_5$ effective potential, but
now in the warped context, to build realistic warped models of the Higgs and 
electroweak symmetry breaking. The central considerations needed to 
achieve realism are worked out in Ref. \cite{bulkrs}. Combining these with 
the Higgs realized as an $A_5$ was done in Ref. \cite{comphiggs2}.


\section{Last Words}

Well, I have just gone through the simplest examples of many of the 
interesting mechanisms connected with extra dimensions. They can appear 
together in interesting combinations. When this happens there is 
greater complexity and the probability for 
error is multiplied. I have found that in the arena of  
warped compactifications, the qualitative insight gained from the AdS/CFT 
connection 
between such compactifications and strongly coupled 4D dynamics, has 
saved me time and time again from errors. It's like "checking units" as an 
undergraduate, in principle it's not necessary, but in practice 
indispensible. But while checking units is dull, the AdS/CFT
connection is pure magic!
It also allows you to design more efficient methods of analysing your models 
quantitatively.
You can learn to think in 
terms of this connection by starting with Refs. \cite{maldacena} 
\cite{rscft}. 

 
\section*{Acknowledgements}

The author is supported by NSF grant P420-D36-2043-4350. He is grateful 
to Dmitry Belyaev for assistance with the figures and equations, and for 
catching several errors.

\end{document}